\documentclass[12pt,epsfig]{article}

\usepackage{epsfig}
\def\beq {\begin{equation}}
\def\eeq {\end{equation}}
\def\nn{\nonumber}

\def\fun#1#2{\lower3.6pt\vbox{\baselineskip0pt\lineskip.9pt
\ialign{$\mathsurround=0pt#1\hfil##\hfil$\crcr#2\crcr\sim\crcr}}}

\begin{document}

\title{Description of composite systems in the spectral integration
technique: the gauge invariance and analyticity constraints for the
radiative decay amplitudes}
\author{V. V. Anisovich  and M. A. Matveev\\ PNPI}
\date{}
\maketitle

\begin{abstract}

The constraints followed from gauge invariance and analyticity are
considered for the amplitudes of radiative transitions of composite
systems when composite  systems are treated in terms of spectral
integrals.  We discuss gauge-invariant amplitudes for the transitions
$S\to\gamma S$ and $V\to\gamma S$ with scalar $S$ and vector
$V$ mesons being two-particle composite systems of scalar (or
pseudoscalar) constituents, and we demonstrate the mechanism of
cancellation of false kinematical singularities. Furthermore, we
explain how to generalize the performed consideration for
quark-antiquark systems, in particular, for  the reaction
$\phi (1020) \to \gamma f_0(980)$. Here we also consider in more
detail the quark-model non-relativistic approach for this reaction.

\end{abstract}

\section{Introduction}

Presently there exists rich information for
radiative decays of  mesons with masses at  1000--1800~MeV,
and one may expect the appearance of more data in future,
see e.g. \cite{L3,Klempt,kloe,barnes} and references therein.
The data on partial widths of radiative decays provide us with an
important knowledge about the quark-gluon structure of hadrons.
However, to suffice this expectation and  avoid misleading
conclusions one needs to work with an adequate technique for the
description of radiative processes involving composite systems.

To describe the low-lying hadronic states, namely,  $S$-wave mesons
of the 36-plet and baryons of the 56-plet in terms of the
SU(6)-symmetry, the non-relativistic quark-model
approach  is an appropriate
technique. The investigation of radiative decays carried out
decades ago played crucial role in the establishing of quark model, by
operating with constituent quark as a universal object for mesons and
baryons \cite{AAA}.
However, for higher states, e.g. for the $P$-wave
quark-antiquark states, relativistic effects become important.
Nevertheless, till now the use of non-relativistic
formulae for radiative decays of  mesons with masses about
1000--1500~MeV  is  rather common practice.

So there is a necessity to use relativistic technique
for the description of radiative decays. The scheme of relativistic
description of the composite system interacting with
electromagnetic field was  suggested in  \cite{deut,epja,ff,rad-yf}.
Within this approach, form factors of composite systems are represented
as double spectral integrals over the masses of composite systems. The
spectral integration technique is a direct generalization of
non-relativistic quantum-mechanical approximation, and the processes
considered within this technique are
time-ordered, like in quantum mechanics.
The energy in the intermediate state is not conserved
but the particles are mass-on-shell. In this method, vertex functions
of the transitions $composite\ system \to constituents$ are defined by
the scattering amplitude of constituents.

The scheme applied to the
composite-system form factors in \cite{deut} is
as follows: partial scattering amplitude $A^{(J)} (s)$
($s$ is total energy squared of the scattered constituents
and $J$ is the total angular momentum) is considered within dispersion
$N/D$ representation \cite{N/D}. In this technique
the amplitude is represented as a sum of dispersion $N/D$ loop diagrams
shown in Fig. 1a. In case when the $D$-function does not have
CDD-poles \cite{CDD}, partial amplitude of the $J$-state reads:
\begin{eqnarray}
A^{(J)}\ =\ Q^{(J)}_{\mu_1\ldots\mu_J}(p_{\perp})\,
\frac{N_J(s)}{1-B_J(s)}\,Q^{(J)}_{\mu_1\ldots\mu_J}(p'_{\perp})\ ,
\label{1.1}
\end{eqnarray}
where $Q^{(J)}_{\mu_1\ldots\mu_J}(p_{\perp})$ is covariant
angular momentum-$J$
operator which depends on relative momentum of particles 1 and 2:
$p_{\perp\,\mu} =g^{\perp}_{\mu\nu}p_\nu $;  here $p=p_1-p_2$, and
$g_{\mu\nu}^{\perp}$ is the metric tensor
which works in the space orthogonal to  $P=p_1+p_2$:
$g_{\mu\nu}^{\perp}= g_{\mu\nu} -P_{\mu}P_{\nu}/P^2$
(for equal particle masses 1 and 2, one has  $p_\perp =p$).
If the scattered particles are spinless and
 $J=L$, where $L$ is the orbital momentum of the system,
we replace
$Q^{(J)}_{\mu_1\ldots\mu_J}(p_{\perp})
\to X^{(L)}_{\mu_1\ldots\mu_L}(p_{\perp})$, where for the
lowest waves, according to \cite{operator}, the
operators $X^{(L)}_{\mu_1\ldots\mu_L}(p_{\perp})$
are determined as follows:
\begin{eqnarray} X^{(0)}(p_{\perp})\ =\ 1\ ,\qquad
X^{(1)}_{\mu}(p_\perp)\ =\ p_{\perp\,\mu}\ ,\qquad
X^{(2)}_{\mu_1\mu_2}(p_\perp)\ =\
\frac32 \left(p_{\perp\,\mu_1}p_{\perp\,\mu_2}-\frac13\,p^2_\perp
g^\perp_{\mu_1\mu_2}\right)\ .
\end{eqnarray}
When the particles 1 and 2 are fermions, like quarks or nucleons, then
the operators $Q^{(J)}_{\mu_1\ldots\mu_J}$ are constructed by using
$\gamma$-matrices, see
\cite{operator} for details.

If there is a bound state with mass
$M$ in the partial wave , the sum of diagrams shown in Fig. 1a
creates a pole at $s=M^2$, see Fig. 1b, and the vertex of the
pole diagram determines the wave function of the bound state.

The form factor of bound state can be defined by the process of Fig. 2a
when the photon is emitted by interacting constituents. The
amplitudes of the initial state (interaction block before the photon
emission) and final one (that after photon emission) contain the poles
$s=M^2$ and $s'=M'^2$, see Fig. 2b,
so the two-pole amplitude defines  form factor of the composite system:
the form factor is the residue in these poles, it is shown separately
in Fig. 2c.

To be more understandable in explaining the form factor
calculus within gauge invariance and analyticity
constraints, we use a simplified variant of the $N$-function, with
separable forces. The hypothesis of separability of the interaction
blocks can be successfully applied to the realistic
description of  composite systems such as the
deuteron is \cite{deut}. Here we use this hypothesis to
simplify rather cumbersome presentation, and this simplification does
not influence the principal statements.  Correspondingly, we use
separable interaction with
$N_J(s)\to g^2_J(s)$.  Then the amplitude of Fig. 2a for the
emission of photon by the two-particle system with total
angular momenta $J$
and $J'$ in initial and final states, respectively, reads as follows:
\begin{eqnarray}
A^{(J\to J')}_\alpha \ =\ Q^{(J)}_{\mu_1\ldots\mu_J}(p_{\perp})\,
\frac{g_J(s)}{1-B_J(s)}\,
\Gamma^{(J\to J')}_{\mu_1\ldots\mu_J\,\alpha\,\nu_1\ldots\nu_{J'}}(P,P';q)\,
\frac{g_{J'}(s')}{1-B_{J'}(s')}
Q^{(J')}_{\nu_1\ldots\nu_{J'}}(p'_{\perp})\ ,
\label{1.3}
\end{eqnarray}
where
$\Gamma^{(\rm J\to\rm
J')}_{\mu_1\ldots\mu_J,\,\alpha,\,\nu_1\ldots\nu_{J'}}$ is the
three-point interaction amplitude at
 $P^2=s$ and $P'^2=s'$. The amplitude $A_\alpha^{(J\to J')}$ is
represented by Fig. 3a as a chain of loop diagrams; the three-point
interaction amplitude is
depicted in Fig. 3a in the middle of the chain of
loop diagrams.

The loop diagram $B_J(s)$, under the ansatz of separable interaction,
is equal to:
\begin{eqnarray}
B_J(s)\ =\ \int\limits^{\infty}_{(m_1+m_2)^2}\,
\frac{d\,\tilde s}{\pi}\,\frac{g^2_J(\tilde s)}{\tilde s-s-i0}\,
\rho_J(\tilde s)\ ,
\end{eqnarray}
where $m_1$ and $m_2$ are the masses of scattered
particles and $\rho_J(s)$ is the phase space  in the
state with total angular momentum  $J$. For scalar
constituent particles and total angular momentum $J=0$,
the phase volume is defined as follows:
\begin{eqnarray}
\label{rho.phi}
&&\rho(s)\ =\ \int d\,\Phi_2(P;p_1,p_2)\ ,
\\
&&d\Phi_2(P;p_1,p_2)\ =\ \frac12\,\frac{d^3p_1}{2p_{10}(2\pi)^3}\,
\frac{d^3p_2}{2p_{20}(2\pi)^3}\,(2\pi)^4\,\delta^{(4)}(P-p_1-p_2)\ ,
\nonumber
\end{eqnarray}
where we have redenoted $\rho_0(s)\to \rho(s)$.
For $J\ne 0$, the convolution of the operators\\
$Q^{(J)}_{\mu_1\ldots\mu_J}(p_{\perp})
Q^{(J)}_{\mu_1\ldots\mu_J}(p_{\perp})$ should be inserted in the
right-hand side of (\ref{rho.phi}) for the calculation of $\rho _J(s)$.

The $D$-function zeros  at $s=M^2$ and $s'=M'^2$ correspond to
the existence of bound states:
\begin{eqnarray}
1-B_J(M^2)\ =\ 0\ , \qquad 1-B_{J'}(M'^2)\ =\ 0\ .
\end{eqnarray}
In this way, the $D$-functions near the pole read:
\begin{eqnarray}
1-B_J(s)\ \simeq \frac{dB_J(M^2)}{ds}(M^2-s)\ , \qquad
1-B_{J'}(s')\ \simeq\
\frac{dB_{J'}(M'^2)}{ds'}(M'^2-s')\ .
\end{eqnarray}
Therefore, the amplitude  (\ref{1.3}) near the poles takes the form:
\begin{eqnarray}
A^{(J\to J')}_\alpha\ \simeq\ Q^{(J)}_{\mu_1\ldots\mu_J}(p_{\perp})\,
\frac{G_J(s)}{M^2-s}\,    \frac{\Gamma^{(J\to
J')}_{\mu_1\ldots\mu_J\,\alpha\,\nu_1\ldots\nu_{J'}}(P,P';q)}
{\sqrt{\frac{dB_J(M^2)}{ds}\frac{dB_{J'}(M'^2)}{ds'}}}\,
\frac{G_{J'}(s')}{M'^2-s'}
Q^{(J')}_{\nu_1\ldots\nu_{J'}}(p'_{\perp})\ ,
\label{1.6}
\end{eqnarray}
where $G_J(s)=g_J(s)/\sqrt{dB_J(M^2)/ds}$ .

One can introduce the wave functions as follows:
\begin{eqnarray}
\psi_J(s)=
\frac{G_J(s)}{M^2-s}\, , \qquad
\psi_{J'}(s')=
\frac{G_{J'}(s')}{M'^2-s'}\, .
\label{1.6a}
\end{eqnarray}
The radiative transition amplitude {\it meson-$J$ $\to$
meson-$J'$} is determined by residues in the poles
$s=M^2$ and $s'=M'^2$:
\begin{eqnarray}
\label{1.6b}
\Gamma^{(meson-J\, \to \,
meson-J')}_{\mu_1\ldots\mu_J\,\alpha\,\nu_1\ldots\nu_{J'}}(P,P';q)=
\frac{\left[\Gamma^{(J\to
J')}_{\mu_1\ldots\mu_J\,\alpha\,\nu_1\ldots\nu_{J'}}(P,P';q)
\right]_{s=M^2,\,\,s'=M'^2 }}{\sqrt{dB_J(M^2)/ds
\,\, dB_{J'}(M'^2)/ds'}} \, .
\end{eqnarray}
In this way, the magnitudes $M^2$ and $M'^2$
are fixed in (\ref{1.6a}). This means that we should discriminate
between analytical properties of the amplitude of photon emission by
unbound particles and those of radiative transition amplitude
$meson$-$J\to meson$-$J'$. Analytical properties of the
amplitude as function of $s$ and $s'$ for the
emission of photon by unbound particles are determines by all diagrams
shown in Fig. 3.

By studying analytical amplitude one should take into account
that moment operators, which give the  spin dependence of the form
factor  $\Gamma^{(J\to
J')}_{\mu_1\ldots\mu_J\,\alpha\,\nu_1\ldots\nu_{J'}}(P,P';q)$, may have
false kinematical singularities. In the whole amplitude of Fig. 3
just these singularities should cancel each other.

This paper is devoted to the problem of cancellation of false
singularities: we use as an example the processes
when  $J=J'=0$ (Section 2) and $J=1$, $J'=0$
(Section 3). We deal with scalar (or
pseudoscalar) constituent particles with equal masses,
$m_1=m_2=m$: this does not affect generality
but simplify cumbersome calculations. Furthermore, in Section 4,
we present the generalization for  quark constituents.

In Section 2, the transition $(J=0) \to (J'=0)$
is considered;
we denote this transition as $S\to\gamma S$. The
three-point amplitude for this transition $\Gamma^{(0\to
0)}_{\alpha}(P,P';q)$ can be expanded in respect to two spin operators:
the transverse one,
$\left(P_\perp+P'_\perp\right)_\alpha =
2\left(P_\alpha-q_\alpha(Pq)/q^2 \right)$,
and longitudinal one,
$q_\alpha$.
Transverse spin operator contains kinematical singularity $1/q^2$,
which should be cancelled in the whole amplitude.

The necessity to use transverse and longitudinal spin operators,
which are orthogonal to each other
$\left(P_\perp+P'_\perp\right)_\alpha\,q_\alpha = 0$,
is dictated by the specifics of the
spectral-representation method for form factors of the
$S\to\gamma S$ transition. Vertex function of the photon emission is
expanded in respect to independent and orthogonal operators
$\left(P_\perp+P'_\perp\right)_\alpha$ and $q_\alpha $ as follows:
\begin{eqnarray}
\Gamma^{(0\to 0)}_{\alpha}(P,P';q)\ =\
\left(P_\perp+P'_\perp\right)_\alpha\, F^{(0\to0)}_T(s,s',q^2)
+q_\alpha\, F^{(0\to0)}_L(s,s',q^2)\ .
\label{1.20}
\end{eqnarray}
Then, $F^{(0\to0)}_T(s,s',q^2)$ and $F^{(0\to0)}_L(s,s',q^2)$ are
defined by dispersion integrals over $s$ and
$s'$, which are in due course determined by  transverse and
longitudinal components of the triangle diagram only.

The mechanism of singularity cancellation in $\Gamma^{(0\to
0)}_{\alpha}(P,P';q)$ was considered in \cite{deut}
for the variant of  $F_T^{(0\to 0)}(s,s',q^2)$
being double spectral integral without subtraction terms.
However, as was shown in \cite{deut}, one cannot deal without
subtraction terms at all: subtraction terms in
$F_L^{(0\to 0)}(s,s',q^2)$  are important for cancelling
false singularities in (\ref{1.20}). In Section 2,
we consider more general case of $\Gamma^{(0\to 0)}_{\alpha}(P,P';q)$
with  subtraction term in
$F_T^{(0\to 0)}(s,s',q^2)$: such a variant serves us as a guide to
consider transition form factors with angular momenta $J=1$, $J'=0$.

The vertex function for the transition $(J=1)\to(J'=0)$,
or $V\to\gamma S$, is considered in Section 3. The spin structure of
such a vertex,  $\Gamma^{(1\to 0)}_{\mu\alpha}(P,P';q)$, is determined
by two independent tensors, one of them can be chosen as metric tensor
operating in the two-dimensional space and being orthogonal to
$P$ and $q$:
\begin{eqnarray}
g^{\perp\perp}_{\mu\alpha}\ =\
g_{\mu\alpha }+\frac{q^2}{(Pq)^2-P^2q^2}\,P_\mu P_\alpha
+\frac{P^2}{(Pq)^2-P^2q^2}\,q_{\mu}q_\alpha
-\frac{(Pq)}{(Pq)^2-P^2q^2}\,(P_{\mu}q_\alpha +q_{\mu} P_\alpha)\ ,
\label{1.11}
\end{eqnarray}
while the second tensor is defined as
\begin{eqnarray}
\label{1.12}
4\,L_{\mu\alpha}\ =&&\
\frac{q^2}{(Pq)^2-P^2q^2}\,P_{\mu}P_\alpha
+\frac{P^2}{(Pq)^2-P^2q^2}\,q_{\mu}q_\alpha
\\
\nonumber
&&-\frac{(Pq)}{(Pq)^2-P^2q^2}\,P_{\mu}q_\alpha
-\frac{P^2q^2}{[(Pq)^2-P^2q^2]\,(Pq)}\,q_{\mu}P_\alpha\ .
\end{eqnarray}
These two tensors satisfy gauge invariance requirements:
\begin{eqnarray}
P_\mu\, g^{\perp\perp}_{\mu\alpha}\ =\ 0\, ,\qquad
g^{\perp\perp}_{\mu\alpha}\, q_\alpha\ =\ 0\, ,\qquad
P_\mu\, L_{\mu\alpha}\ =\ 0\, ,\qquad
L_{\mu\alpha}\, q_\alpha\ =\ 0\ ,
\label{1.13}
\end{eqnarray}
and $L_{\mu\alpha}$ is constructed to be orthogonal to
$g^{\perp\perp}_{\mu\alpha}$:
\begin{eqnarray}
L_{\mu\alpha}\, g^{\perp\perp}_{\mu\alpha}\ =\ 0\ .
\label{1.14}
\end{eqnarray}
Vertex function for $V\to\gamma S$ is defined by two form factors
which correspond to operators in the form  (\ref{1.11})
and (\ref{1.12}):
\begin{eqnarray}
\label{1.15}
\Gamma^{(1\to 0)}_{\mu\alpha}(P,P';q)\ =\
g^{\perp\perp}_{\mu\alpha}\, F^{(1\to 0)}_T(s,s',q^2)
+L_{\mu\alpha}\, F^{(1\to 0)}_L(s,s',q^2)\ .
\end{eqnarray}
The operators written in  (\ref{1.11}) and (\ref{1.12}) are singular:
they contain poles at $(Pq)^2-P^2q^2=0$ and $(Pq)=0$.
In (\ref{1.15}) these
singularities must be compensated by zeros of the form factors
$F^{(1\to0)}_T(s,s',q^2)$ and $F^{(1\to0)}_L(s,s',q^2)$; corresponding
mechanism is considered in detail in Section 3.
We demonstrate that to compensate false singularities in the
reactions $V\to\gamma S$ the subtraction terms play an important role;
from this point of view the compensation mechanism in the
amplitudes $S\to\gamma S$ and $V\to\gamma S$ is similar.

At $q^2=0$, the only form factor $F^{(1\to 0)}_T(s,s',q^2)$
determines the transition amplitude $V\to\gamma S$. In Section 3, we
discuss in detail the problem of unambigous determination of the
amplitude in terms of the spectral integration technique.
The matter is that at $q^2\to0$ one of independent operators of
(\ref{1.11}) and (\ref{1.12}) become nilpotent one.
Indeed,
\begin{eqnarray}
\label{1.16}
&&g^{\perp\perp}_{\mu\alpha}(0)\ =\
g_{\mu\alpha }+\frac{4\,s}{(s-s')^2}\,q_{\mu}q_\alpha
-\frac{2}{s-s'}\,(P_{\mu}q_\alpha +q_{\mu} P_\alpha)\ ,
\\
\nn
&&L_{\mu\alpha}(0)\ =\
\frac{s}{(s-s')^2}\,q_{\mu}q_\alpha
-\frac{1}{2(s-s')}\,P_{\mu}q_\alpha\ ,
\end{eqnarray}
and the second operator has a zero norm:
\begin{eqnarray}
L_{\mu\alpha}(0)\, L_{\mu\alpha}(0)\ =\ 0\ .
\label{1.17}
\end{eqnarray}
Due to Eqs.
(\ref{1.14}) and (\ref{1.17}) any combination of
$g^{\perp\perp}_{\mu\alpha}(0)$ and $L_{\mu\alpha}(0)$,
\begin{eqnarray}
g^{\perp\perp}_{\mu\alpha}(0)+C(s,s')\,L_{\mu\alpha}(0)\ ,
\label{1.18}
\end{eqnarray}
can be equally used to define the transverse form factor
$F_T(s,s',0)$. In Section 3, this propery is demonstrated
directly by considering two sets of
operators, the first one given by (\ref{1.16})
and the second set appeared after the substitution
as follows:
\begin{eqnarray}
g^{\perp\perp}_{\mu\alpha}(0)\to
g^{\perp\perp}_{\mu\alpha}(0)+4\, L_{\mu\alpha}(0)\ =\
g_{\mu\alpha}-\frac{2}{s-s'}\,q_\mu P_\alpha\ .
\label{1.19}
\end{eqnarray}
Such a choice of operators for
the illustration of (\ref{1.18}) is motivated by
the existing statements (for example,  see \cite{acha})
that only the form of (\ref{1.19}) provide us with a correct
operator expansion of  transition amplitude for
$V\to\gamma(q^2=0)\,S$. However, our consideration performed in Section
3 proves directly that it is not so.

Concerning analytical properties
of amplitudes in the transitions $V\to \gamma S$, with emission of
photon ($q^2=0$), they are as follows.

The whole  amplitude of the transition of unbound particles
is a sum of
three contributions shown in Fig. 3a,b,c:
\begin{eqnarray}
\frac{g_1(s)}{1-B_1(s)}
\left [g_{\mu\alpha}^{\perp\perp}(0)\, T(s,s',0)+
L_{\mu\alpha}(0)\, L(s,s',0)\right]
\frac{g_0(s')}{1-B_0(s')}\ .
\label{Eq21}
\end{eqnarray}
The analyticity of this amplitude requires for transverse component
$T(s,s',0)$ to have zero of the first order at $s=s'$, namely,
$T(s,s,0)=0$, while the combination $L(s,s',0)+4T(s,s',0)$ should have
zero of the second order.

Transverse amplitude for the transition of bound vector state to bound
scalar state, $F_T^{(1\to 0)}(s=M^2,\, s'=M'^2,\, 0)$, which is defined
by the diagram of Fig. 2c, is not indebted to be zero at
$\omega=M-M'=0$.

In Section 4, we show the way of generalization of
our results for constituent quarks.
 We present the transition form factor for
$\phi (1020) \to \gamma f_0(980)$ both in the form of relativistic
spectral integral \cite{epja} and non-relativistic quark model approach.

Direct form factor calculations performed for the transition
$\phi(1020) \to \gamma f_0(980)$ both in relativistic and
non-relativistic approximations demonstrate the absence of zero at
$\omega\to 0$. This is in accordance with general result obtained in
the analysis of the whole amplitude (\ref{Eq21}).

In Conclusion we summarise briefly the results focussing our attention
to advantages of the spectral integration technique and the
problems which this method faces on.

\section{Interaction of scalar two-particle composite system
in the state $J^{P}=0^{+}$ with electromagnetic field: $S\to
\gamma S$ }

Diagrammatical representation of the interaction amplitude in terms of
the dispersion-relation graphs is shown in Fig. 3. These diagrams are
obtained by coupling the photon to all the constituents (both
internal and external) in the diagrams related to the constituent
scattering amplitude. Besides, there is an uncoupled
diagram of Fig. 3d corresponding to non-interacting constituents.

We assume that two constituents may form the bound state, and the form
factor of composite system is determined  by  residues in the amplitude
poles, see Fig. 2b.

For the sake of simplicity, the constituent masses are put to be
equal to each other, $m_1=m_2=m$, though constituents are not
identical; in this way we do not symmetrize the states
of  constituents.

\subsection{Diagrams of the Fig. 3a type}

Consider the sum of the diagrams shown in Fig. 3a, when constituents
interact in the $S$-wave state both before and after the emission of
photon. The sum of diagrams of such a type is written as:
\beq
A^{(0\to0)}_\alpha(P,P';q)\ =\ \frac{g_0(s)}{1-B_0(s)}\
\Gamma^{(0\to0)}_\alpha(P,P';q)\, \frac{g_0(s')}{1-B_0(s')}\ ,
\eeq where
$\Gamma^{(0\to0)}_\alpha$ is the three-point function shown separately
in Fig. 4a; its representation through form factors
$F^{(0\to0)}_T(s,s',q^2)$ and $F^{(0\to0)}_L(s,s',q^2)$ is given in
(\ref{1.20}).

The dispersion representation of the triangle graph can be found
in \cite{deut,epja,ff,rad-yf}. Here we briefly repeat the scheme
keeping in mind to apply it not only to three-point
diagrams but also to the two-point ones describing the photon emission
by constituents in the initial and final states too, see Fig.
4b,c.

Let us start with the Feynman
expression for the triangle diagram
$\Gamma^{(0\to0)}_\alpha(P,P';q)$:
\beq
\int\frac{d^4k_1}{i(2\pi)^4}\
\frac{g_0(\tilde P^2;k^2_1,k^2_2) (k_1+k'_1)_\alpha
g_0(\tilde P'^2;k'^2_1,k^2_2)}{(m^2-k^2_1)(m^2-k^2_2)(m^2-k'^2_1)}\ .
\label{intf}
\eeq
The following steps are necessary to write down the dispersion integral
starting from this amplitude:

(i) We should calculate the double discontinuity of the Feynman
diagram (\ref{intf}), with fixed energy squared of initial and
final states, $\tilde P^2 =\tilde s$ and $\tilde P'^2 =\tilde s'$.
This implies the substitution of operators
in the intermediate states as follows:
\begin{eqnarray}
&& (m^2-k^2_1)^{-1}(m^2-k^2_2)^{-1}\to\
\theta(k_{10})\delta(k^2_1-m^2)
\theta(k_{20})\delta(k^2_2-m^2)\
\nonumber
\\
&& (m^2-k'^2_1)^{-1}\to\,  \theta(k'_{10})\,
\delta(k'^2_1-m^2)
\end{eqnarray}
as well as integration over three-particle phase space in both channels
at fixed $\tilde s$ and $\tilde s'$:
\begin{eqnarray}
&&\frac{d^4k_1}{(2\pi)^4}\,
\delta(m^2-k^2_1)\,\delta(m^2-k'^2_1)\,\delta(m^2-k^2_2)\
\longrightarrow\
d\Phi_{tr}\left(\tilde P,\tilde P';k_1,k'_1,k_2\right)\ ,
\nonumber
\\
&& d\Phi_{tr}\left(\tilde P,\tilde P';k_1,k'_1,k_2\right)
=\ d\Phi_2 (\tilde P;k_1,k_2)d\Phi_2(\tilde P';k'_1,k'_2)
2k_{20}(2\pi)^3 \delta^3 ({\bf k}_2-{\bf k}'_2)\ ,
\label{tr.c}
\end{eqnarray}
The constituents in the
intermediate state are mass-on-shell. The double discontinuity is
calculated at  $\tilde q=\tilde P-\tilde P'\neq q$ but $\tilde
q^2=q^2$.

(ii) The vertex functions are to be replaced as
\beq
g_0(\tilde
P^2;k^2_1,k^2_2)\,g_0(\tilde P'^2;k'^2_1,k^2_2)\ \longrightarrow\
g_0(\tilde s)\, g_0(\tilde s')\ ,
\eeq
that actualize the treatment of composite system as a true
two-particle state.

(iii)  The invariant part of the triangle diagram
should  be singled out by
expanding the spin factor $(k_1+k'_1)_\alpha$ in the vectors
$(\tilde P+\tilde P')_{\perp}$
and $\tilde q=\tilde P-\tilde P'$:
\beq
(k_1+k'_1)_\alpha\ =\ \alpha(\tilde s,\tilde s',q^2)\,
\left(\tilde P_\alpha+\tilde P'_\alpha-\frac{\tilde s-\tilde s'}{q^2}\,\tilde
q_\alpha\right)+\beta(\tilde s,\tilde s',q^2)\tilde q_\alpha\ .
\label{coef1}
\eeq
The coefficients $\alpha(\tilde s,\tilde s',q^2)$ and
$\beta(\tilde s,\tilde s',q^2)$ are given below, in Eq. (\ref{32}).
As a result, we have the following expressions for double
discontinuities (double spectral densities) for the triangle diagram:
\begin{eqnarray}
&& disc_{\tilde s}disc_{\tilde s'}
F^{(0\to0)}_T(\tilde s,\tilde s',q^2)
=\alpha(\tilde s,\tilde s',q^2)g_0(\tilde s)g_0(\tilde s')
\,d\Phi_{tr}(\tilde P,\tilde P';k_1,k_2,k'_1),
\nonumber
\\
&& disc_{\tilde s}disc_{\tilde s'}
F^{(0\to0)}_L(\tilde s,\tilde s',q^2)
=\beta(\tilde s,\tilde s',q^2)g_0(\tilde s)g_0(\tilde s')
\,d\Phi_{tr}(\tilde P,\tilde P';k_1,k_2,k'_1)\ .
\end{eqnarray}
The form factor $F^{(0\to0)}_i(s,s',q^2)$ is determined by its
spectral density as follows:
\beq
F_i^{(0\to0)}(s,s',q^2)\ =\ f^{(0\to0)}_i(s,s',q^2)+\int\limits^\infty_{4m^2}
\frac{d\tilde s}\pi\frac{d\tilde s'}\pi\,
\frac{disc_{\tilde s}disc_{\tilde s'}
F^{(0\to0)}_i(\tilde s,\tilde s',q^2)}{(\tilde s-s-i0)\,(\tilde s'-s'-i0)}\ ,
\label{ff1}
\eeq
where $f^{(0\to0)}_i(s,s',q^2)$ are the subtraction terms with zero
double spectral density. Within the approach, where partial amplitude
is described by a set of dispersion diagrams of Fig. 1a, the
subtraction term   $f^{(0\to0)}_i(s,s',q^2)$ is an arbitrary function
determined by diagrams where photon interacts with other particles, not
constituents, for example, with mesons which determine the forces
between constituents.

The expansion coefficients (\ref{coef1}) are calculated under the
orthogonality requirements. Indeed, by projecting (\ref{coef1}) onto
$\left(\tilde P_\alpha+\tilde P'_\alpha
-\tilde q_\alpha(\tilde s-\tilde s')/q^2\right)$ and
$\tilde q_\alpha$, we obtain the equations for
$\alpha(\tilde s,\tilde s',q^2)$ and
$\beta(\tilde s,\tilde s',q^2)$.
We have:
\begin{eqnarray}
&& \alpha(\tilde s,\tilde s',q^2)\ =\
-\,\frac{ q^2(\tilde s+\tilde s'-q^2)}
{\lambda(\tilde s,\tilde s',q^2)}\ ,
\nn
\\
&& \beta(\tilde s,\tilde s',q^2)\ =\ 0\ ,
\nn
\\
&&\lambda(\tilde s,\tilde s',q^2)\ =\
-2 q^2 (\tilde s+\tilde s')+q^4+(\tilde s-\tilde s')^2 \ .
\label{32}
\end{eqnarray}
Here we took into account that $k^2_i=m^2$ and $(k_1-k'_1)^2=q^2 $.
Therefore, $F^{(0\to0)}_L(s,s',q^2)$ has zero double spectral density,
and it is defined by the subtraction term only:
\beq
F^{(0\to0)}_L(s,s',q^2)\ =\ f^{(0\to0)}_L(s,s',q^2)\ .
\eeq
For $F^{(0\to0)}_T(s,s',q^2)$, after integrating in (\ref{tr.c})
over the momenta $k_1$, $k'_1$ and $k_2$ at fixed $s$ and $s'$,
we obtain the following equation:
\beq
F^{(0\to0)}_T(s,s',q^2)=f^{(0\to0)}_T(s,s',q^2)+\int\limits^{\infty}_{4m^2}
\frac{d\tilde s \,d\tilde s'}{\pi^2}
\,\frac{g_0(\tilde s)}{\tilde s -s}
\,\frac{g_0(\tilde s')}{\tilde s' -s'}
\,\frac{\Theta\left(-\tilde s\,\tilde s'\,q^2-m^2\lambda(\tilde s,\tilde s',q^2)\right)}
{16\sqrt{\lambda(\tilde s,\tilde s',q^2)}}\,
\alpha(\tilde s,\tilde s',q^2)\ .
\label{ff1.sp}
\eeq
Here the $\Theta$-function is defined as follows: $\Theta(X)=1$ at
$X\ge 0$ and $\Theta(X)=0$ at $X<0$.

Now let us come back to the requirements the amplitude should obey.
First, as was said above, it must be analytical function, that is,
kinematical singularities should be absent. Concerning
the $q^2=0$ singularity of the considered amplitude,
the term we should  care for  is:
\beq
-\,\frac{s-s'}{q^2}\, F^{(0\to0)}_T(s,s',q^2)\, q_\alpha \ .
\eeq
Let us calculate the form factor in the limit $q^2 \to 0$.
To this aim let us introduce new variables in (\ref{ff1.sp}):
\beq
\label{new-variable}
\sigma=\frac12(\tilde s+\tilde s')\ ; \quad \Delta=\tilde s-\tilde s',
\quad Q^2=-q^2\ ,
\eeq
and then consider the case of interest, $Q^2\to 0$. The form factor
formula reads:
\beq
F^{(0\to0)}_T(s,s',q^2)=f^{(0\to0)}_T(s,s',q^2)+\int\limits^\infty_{4m^2}
\frac{d\sigma}\pi\,\frac{g^2_0(\sigma)}{(\sigma-s)(\sigma-s')}
\int\limits^b_{-b}d\Delta\,\frac{\alpha(\sigma,\Delta,Q^2)}
{16\pi\sqrt{\Delta^2+4\sigma Q^2}}\ ,
\eeq
where
\beq
b=\frac{Q}m\ \sqrt{\sigma(\sigma-4m^2)}\ , \quad
\alpha(\sigma,\Delta,Q^2)\ =\
\frac{2\sigma\,Q^2}{\Delta^2+4\sigma Q^2}\ .
\label{b}
\eeq
As a result we have:
\beq
F^{(0\to0)}_T(s,s',0)\ =\ f^{(0\to0)}_T(s,s',0)+\frac{B_0(s)-B_0(s')}{s-s'}\ ,
\eeq
where $B_0(s)$ is the loop diagram:
\begin{eqnarray}
B_0(s)\ =\int \limits^\infty_{4m^2}\frac{d\tilde s}{\pi}\,
\frac{g^2_0(\tilde s)}{\tilde s-s}\rho(\tilde s)\ , \qquad
\rho(s)\ =\ \frac{1}{16\pi}\sqrt{\frac{s-4m^2}{s}}\ .
\label{rho}
\end{eqnarray}
In (\ref{rho}) the index $J$ for the phase volume with
$J=0$ is omitted.

In the limit $q^2\to 0$ the amplitude takes the form:
\begin{eqnarray}
&&\Gamma^{(0\to0)}_\alpha(P,P';q^2\to 0)\  =\
\nn
\\
&&\left(P_\alpha+P'_\alpha-\frac{s-s'}{q^2}\,q_\alpha\right)
\left[f^{(0\to0)}_T(s,s',q^2\to0)+\frac{B_0(s)-B_0(s')}{s-s'}
\right]
\nn
\\
&&+q_\alpha \, f^{(0\to0)}_L(s,s',q^2\to0)\ .
\label{ampl1t0}
\end{eqnarray}
The amplitude (\ref{ampl1t0}) should not have pole singularity
$1/q^2$: the presence in the right-hand side of (\ref{ampl1t0})
of singular factor $q_\alpha\,(s-s')/q^2$ is an artifact of our
expansion of the amplitude $\Gamma^{(0\to0)}_\alpha$ in the
transverse and longitudinal components. Therefore, the subtraction term
in $f^{(0\to0)}_L(s,s',q^2\to0)$ must contain expressions which
cancel the singularity $1/q^2$.  The calcellation of singular terms
lead to the requirement:
\beq
f^{(0\to0)}_L(s,s',q^2\to0)\ =\
\frac{1}{q^2} \left((s-s')f^{(0\to0)}_T(s,s',0)+B_0(s)-B_0(s')\right)\
.  \label{f_s_2}
\eeq
After the fulfilment of (\ref{f_s_2}), we have for
$\Gamma^{(0\to0)}_\alpha(s,s';0)$:
\beq
\Gamma^{(0\to0)}_\alpha(s,s';0)\ =\ \left(P_\alpha+P'_\alpha\right)
\left(f^{(0\to0)}_T(s,s',0)+\frac{B_0(s)-B_0(s')}{s-s'}\right)\ .
\label{gam.c.1}
\eeq
This formula has been obtained in \cite{deut} for the case of
$f^{(0\to0)}_T\equiv 0$. In this approximation we come to a well-known
Ward identity for the triangle diagram:
$q_\alpha\Gamma^{(0\to0)}_\alpha(s,s',0)=B_0(s)-B_0(s')$.
With non-zero subtraction term, the Ward identity looks as follows:
\beq
q_\alpha\Gamma^{(0\to0)}_\alpha(s,s';0)\ =\
(s-s')f^{(0\to0)}_T(s,s',0)+B_0(s)-B_0(s')\ .
\eeq

\subsection{Diagrams of   Fig. 3b,c  type }

Consider the amplitude for the diagram of Fig.  3b; it reads
as follows:
\beq
A^{(\to0)}_\alpha(P,P';q)\ =\ \Gamma^{(\to0)}_\alpha(P,P';q)
\frac{g_0(s')}{1-B_0(s')}\ .
\label{lt.tot}
\eeq
Here $\Gamma^{(\to0)}_\alpha$ stands for the vertex representing the
emission of photon by the incoming constituent, it is shown
in Fig. 4b.

By singling out the $S$-wave state from the initial state of the
amplitude of Fig. 4b,
$\Gamma^{(\to 0)}_\alpha\to \Gamma^{(S\to 0)}_\alpha$,
we can represent $\Gamma^{(S\to 0)}_\alpha$ as the spectral
integral.
 In this way the amplitude is written as follows:
\beq \Gamma^{(\rm S\to0)}_\alpha (P,P';q)\ =\ \left(P_\alpha+P'_\alpha
-\frac{s-s'}{q^2}q_\alpha\right)
F^{(\rm S\to0)}_T(s,s',q^2)+q_\alpha \, F^{(\rm S\to0)}_L(s,s',q^2)\ .
\label{ampl2}
\eeq
The spectral integrals for $F^{(\rm S\to0)}_T$ and $F^{(\rm S\to0)}_L$
are obtained in the same way as before. Namely,
we project the Feynman of Fig. 4b on
the $S$-wave state by averaging over the phase space of initial
particles, $k_1$ and $k_2$:
\beq
\int\frac{d\Phi_2(P;k_1,k_2)}{\rho(s)}
\, (k_1+k'_1)_\alpha\,\frac{ g_0(\tilde P^2;k'^2_1,k^2_2)}{m^2-k'^2_1}
\ .
\label{ampl2f}
\eeq
The discontinuity of the amplitude  (\ref{ampl2})
is calculated for the
mass-on-shell constituent:
\beq
(m^2-k'^2_1)^{-1}\ \longrightarrow\
\theta(k'_{10})\delta (k'^2_1-m^2) \ ,
\eeq
with the phase space  integration
in the channel  $s'$:
\beq
(2\pi)^32k_{20}\delta^{(3)}({\bf
k}_2-{\bf k}'_2)\, d\Phi_2(P';k'_1,k'_2)\ .
\eeq
Besides, it is
necessary to expand the spin factor $(k_1+k'_1)_\alpha$ in the
vectors $P_\alpha$, $\tilde P'_\alpha$ and  $\tilde q$ with
the constraint $\tilde
q^2=q^2$.  Invariant expansion coefficients, $\alpha(s,\tilde
s',q^2)$ and $\beta(s,\tilde s',q^2)$, are given by Eqs. (\ref{coef1})
and (\ref{32}).

After the
substitution $g_0(\tilde P\,'^2 ;k'^2_1,k^2_2)\ \to\ g_0(\tilde
s\, ')$ and spectral integration, we have the following
representation for  $F^{(\rm S\to0)}_i(s,s',q^2)$:
\begin{eqnarray}
&&F^{(\rm S\to0)}_T(s,s',q^2)\
=\ f^{(\rm S\to0)}_T(s,s',q^2)+\int\limits^\infty_{4m^2}
\frac{d\tilde s'}\pi\,\alpha(s,\tilde s',q^2)\,
\frac{g_0(\tilde s')}{\tilde s'-s'}\,
d\Phi_{tr}(P,\tilde P\,';k_1,k'_1,k_2)\ ,
\nn
\\
&&F^{(\rm S\to0)}_L(s,s',q^2)\
=\ f^{(\rm S\to0)}_L(s,s',q^2)\ .
\end{eqnarray}
Here we took
into account that $\beta(s,\tilde s',q^2)=0$, see
(\ref{32}); the     equation for
$F^{(\rm S\to0)}_L(s,s',q^2)$ has  zero double spectral density,
and it is completely determined by its subtraction term.

For $F^{(\rm S\to0)}_T(s,s',q^2)$, after integrating over
$k_1$, $k'_1$ and  $k_2$ at fixed $s'$, we have the following
expression:
\beq F^{(\rm
S\to0)}_T(s,s',q^2)=f^{(\rm S\to0)}_T(s,s',q^2)
+\int\limits^{\infty}_{4m^2}\frac{d\tilde s'}{\pi}\,
\frac{g_0(\tilde s')}{\tilde s' -s'-i0}\,
\frac{\Theta\left(-s\,\tilde s'\,q^2-m^2\lambda(s,\tilde s',q^2)\right)}
{16\sqrt{\lambda(s,\tilde s',q^2)}}\,
\alpha(s,\tilde s',q^2)\ .
\eeq
For the factor
$F^{(\rm S\to0)}_T(s,s',q^2)$ in the limit $q^2\to
0$, after the same calculations as for previous diagrams,
we obtain:
\beq
F^{(\rm S\to0)}_T(s,s',0)\ =\ f^{(\rm S\to0)}_T(s,s',0)+\frac{g_0(s)}{s-s'}\ .
\eeq
The amplitude (\ref{ampl2}) in the limit $q^2\to 0$ has the form:
\begin{eqnarray}
&&\Gamma^{(\rm S\to0)}_\alpha(P,P';q^2\to 0)\ =\
\nn
\\
&=&\left(P_\alpha+P'_\alpha-\frac{s-s'}{q^2}\,q_\alpha\right)
\left[f^{(\rm S\to0)}_T(s,s',q^2\to0)+\frac{g_0(s)}{s-s'}\right]+
\nn
\\
&&+q_\alpha \, f^{(\rm S\to0)}_L(s,s',q^2\to0)\ ,
\label{ampl2t0}
\end{eqnarray}
and the requirement of absence of singularity $1/q^2$ in the amplitude
$\Gamma^{(\rm S\to0)}_\alpha(P,P';q^2\to 0)$ leads to the formula:
\beq
f^{(\rm S\to0)}_L(s,s',q^2\to0)\ =\ \frac{1}{q^2}
\left((s-s')f^{(\rm S\to0)}_T(s,s',0)+g_0(s)\right)\ .
\label{f.s.lt.2}
\eeq
Taking this condition into account for the vertex
$\Gamma^{(\rm S\to0)}_\alpha(P,P';0)$, one obtains:
\beq
\Gamma^{(\rm S\to0)}_\alpha(P,P';0)\ =\
\left(P_\alpha+P'_\alpha\right)
\left(f^{(\rm S\to0)}_T(s,s',0)+\frac{g_0(s)}{s-s'}\right)\ .
\label{gam.lt.1}
\eeq
The amplitude for diagrams of Fig. 3c-type is treated similarly.
After the $S$-wave is extracted for outgoing constituents, we have:
\beq
A^{(0\to\rm S)}_\alpha(P,P';q)\ =\ \frac{g_0(s)}{1-B_0(s)}\,\,
\Gamma^{(0\to\rm S)}_\alpha(P,P';q)\ ,
\eeq
where
\beq
\Gamma^{(0\to\rm S)}_\alpha (P,P';q)\ =\ \left(P_\alpha+P'_\alpha
-\frac{s-s'}{q^2}q_\alpha\right)
F^{(0\to\rm S)}_T(s,s',q^2)+q_\alpha \, F^{(0\to\rm S)}_L(s,s',q^2)\ ,
\label{ampl3}
\eeq
and
\begin{eqnarray}
&&F^{(0\to\rm S)}_L(s,s',q^2)\ =\ f^{(0\to\rm S)}_L(s,s',q^2)\ ,
\\
&&F^{(0\to\rm S)}_T(s,s',q^2)=f^{(0\to\rm S)}_T(s,s',q^2)
+\int\limits^{\infty}_{4m^2}\frac{d\tilde s}{\pi}\,
\frac{g_0(\tilde s)}{\tilde s -s}\,
\frac{\Theta\left(-\tilde s\,s'\,q^2-m^2\lambda(\tilde s,s',q^2)\right)}
{16\sqrt{\lambda(\tilde s,s',q^2)}}\,
\alpha(\tilde s,s',q^2)\ .
\nn
\end{eqnarray}

In the limit $q^2\to 0$, the amplitude takes the form:
\begin{eqnarray}
&&\Gamma^{(0\to\rm S)}_\alpha(P,P';q^2\to 0)\ =\
\nn
\\
&=&\left(P_\alpha+P'_\alpha-\frac{s-s'}{q^2}\,q_\alpha\right)
\left[f^{(0\to\rm S)}_T(s,s',q^2\to0)-\frac{g_0(s')}{s-s'}\right]+
\nn
\\
&+& q_\alpha \, f^{(0\to\rm S)}_L(s,s',q^2\to0)\ ,
\label{ampl2t0a}
\end{eqnarray}
and the requirement of cancellation of the singularity $1/q^2$
works out the following formula:
\begin{eqnarray}
&&f^{(0\to\rm S)}_L(s,s',q^2\to0)\ =\ \frac{1}{q^2}
\left((s-s')f^{(0\to\rm S)}_T(s,s',0)-g_0(s')\right)\ .
\end{eqnarray}
That results in
\begin{eqnarray}
&&\Gamma^{(0\to\rm S)}_\alpha(P,P';0)\ =\
\left(P_\alpha+P'_\alpha\right)
\left(f^{(0\to\rm S)}_T(s,s',0)-\frac{g_0(s')}{s-s'}\right)\ .
\label{gam.r.1}
\end{eqnarray}

\subsection{Connected diagrams in the limit $q^2\to0$}

First, consider the case when there are no subtraction terms
in the transverse form factor,
$f^{(S\to 0)}_T(s,s',0)=f^{(0\to S)}_T(s,s',0)=0$:
such a variant has
been considered in \cite{deut}.  It is easy to see that the sum of
all connected diagrams of Fig. 3a, 3b and 3c, is equal to zero in this
limit. Indeed , according to (\ref{gam.c.1}),
(\ref{gam.lt.1}) and (\ref{gam.r.1}), it is equal to
\beq
\left(P_\alpha+P'_\alpha\right)\, \left[
\frac{g_0(s)}{s-s'}\,\frac{g_0(s')}{1-B_0(s')}
+\frac{g_0(s)}{1-B_0(s)}\,\frac{B_0(s)-
B_0(s')}{s-s'}\,\frac{g_0(s')}{1-B_0(s')}
-\frac{g_0(s)}{1-B_0(s)}\,\frac{g_0(s')}{s-s'}
\right]
\ =\ 0\ .
\eeq
Generally, when subtraction terms differ from zero,
the amplitude $A^{(conected)}_\alpha(P,P',q^2)$ should also turn into
zero at $q^2\to0$:
\begin{eqnarray}
&&A^{(conected)}_\alpha(P,P',q^2\to0)\ =\
\\
\nn
&&=A^{(\rm S\to0)}_\alpha(P,P',q^2\to0)+A^{(0\to0)}_\alpha(P,P',q^2\to0)
+A^{(0\to\rm S)}_\alpha(P,P',q^2\to0)\ =\ 0\ ,
\end{eqnarray}
that is equivalent to the requirement
(see formulae (\ref{gam.c.1}), (\ref{gam.lt.1}), (\ref{gam.r.1})):
\beq
f^{(0\to0)}_T(s,s',0)\ =\
\frac{B_0(s)-1}{g_0(s)}\,f^{(\rm S\to0)}_T(s,s',0)+f^{(0\to\rm S)}_T(s,s',0)\,\frac{B_0(s')-1}{g_0(s')}\ .
\label{f.c.lt.r}
\eeq
Let us bring our attention to the following. The existence of  bound
state requires $B_0(M^2)=1$; therefore, we have on the basis of
 (\ref{f.c.lt.r}):
\beq
f^{(0\to0)}_T(M^2,M^2,0)=0\ .
\label{f.s.c.0}
\eeq
This means that charge form factor of composite system
$F(q^2)=F^{(0\to0)}_T(M^2,M^2,q^2)$
is determined
at $q^2=0$ by the triangle graph only, without subtraction terms.
Such a property is not surprising:  relying on $B_0(M^2)=1$, we assumed
actually that composite system is true two-particle one, and the
condition for charge form factor of the composite system,
\beq
F(0)=  1 \ ,
\eeq
is the normalization condition for the wave function of this system.

Accounting for (\ref{f.s.c.0}), one can impose more general constraint
\beq f^{(0\to0)}_T(M^2,M^2,q^2)\ =\ 0\ ,  \eeq
that is equivalent to the suggestion that
 charge form factor of composite system is
defined by double spectral integral only. Within this
approximation the form factor of deuteron as two-nucleon system was
calculated in \cite{deut}  as well as form factor of the pion
treated as $q\bar q$-system \cite{ff}.

\section{Transition $V\to\gamma S$}

Consider now  the transition of
vector state to scalar one,  $V\to\gamma S$: this is the transition of
the $P$-wave two-constituent state to the  $S$-wave one. Such a
reaction, as in the previous case, is represented by a set of
diagrams shown in Fig. 3a,b,c.  The vertex function for  $V\to \gamma
S$,  $\Gamma^{(1\to0)}_{\mu\alpha}$, depends on the two spin
indices:  $\mu$ stands for vector state of constituents and $\alpha$,
as before, is related to photon.

In the diagrams shown in Fig. 4a,b,c the following factor carries
spin indices:
\beq
\left(k_1-k_2\right)_{\mu}\left(k_1+k'_1\right)_{\alpha},
\eeq
where $\left(k_1-k_2\right)_{\mu}$ provides the $P$-wave of the initial
state and $\left(k_1+k'_1\right)_{\alpha}$ determines gauge-invariant
vertex {\it photon--constituent}. Let us expand the factor
$\left(k_1-k_2\right)_{\mu}\left(k_1+k'_1\right)_{\alpha}$
in the spin operators. As was said in Introduction,
there exists a freedom in the choice of expansion operators. To reveal
the consequences of this freedom for spectral amplitudes, consider in
parallel two sets of operators. In the first case the operators are as
follows:
\begin{eqnarray}
\rm{Expansion \,\,\, I:}&&
g^{\perp\perp}_{\mu\alpha}\ ,\qquad L_{\mu\alpha}\ ,
\end{eqnarray}
and in the second one:
\begin{eqnarray}
\rm{Expansion \,\,\, II:}&&
\widetilde g^{\perp\perp}_{\mu\alpha}\ =\
g_{\mu\alpha }-\frac{q_{\mu}P_\alpha}{(Pq)}\ =\
g^{\perp\perp}_{\mu\alpha }-4\,L_{\mu\alpha}\ ,\qquad L_{\mu\alpha}\ .
\end{eqnarray}
Recall that the operators
$g^{\perp\perp}_{\mu\alpha}$ and $ L_{\mu\alpha}$ were introduced in
(\ref{1.11}) and (\ref{1.12}).

The convolutions of operators $g^{\perp\perp}_{\mu\alpha}$,
$\widetilde g^{\perp\perp}_{\mu\alpha}$ and $L_{\mu\alpha}$ are
equal to:
\begin{eqnarray}
&&g^{\perp\perp}_{\mu\alpha}\, g^{\perp\perp}_{\mu\alpha}\ =\ 2\ ,
\qquad
L_{\mu\alpha}\, L_{\mu\alpha}\ =\ \frac{q^2\,P^2}{16\, (Pq)^2}\ ,
\qquad
L_{\mu\alpha}\, g^{\perp\perp}_{\mu\alpha}\ =\ 0\ ,
\nn
\\
&&\widetilde g^{\perp\perp}_{\mu\alpha}\, \widetilde g^{\perp\perp}_{\mu\alpha}
\ =\ 2+\frac{q^2\,P^2}{(Pq)^2}\ ,
\qquad
L_{\mu\alpha}\, \widetilde g^{\perp\perp}_{\mu\alpha}\ =\
\frac{q^2\,P^2}{4\, (Pq)^2}\ .
\label{ortnorm}
\end{eqnarray}
We see that the operators from the second set are not orthogonal to one
another but the orthogonality is restored at $q^2\to 0$.
At $q^2=0$ one has:
\begin{eqnarray}
&&g^{\perp\perp}_{\mu\alpha}(0)\, g^{\perp\perp}_{\mu\alpha}(0)\ =\ 2\ ,
\qquad
L_{\mu\alpha}(0)\, L_{\mu\alpha}(0)\ =\ 0\ ,
\qquad
L_{\mu\alpha}(0)\, g^{\perp\perp}_{\mu\alpha}(0)\ =\ 0\ ,
\nn
\\
&&\widetilde g^{\perp\perp}_{\mu\alpha}(0)\, \widetilde g^{\perp\perp}_{\mu\alpha}(0)
\ =\ 2\ ,
\qquad
L_{\mu\alpha}(0)\, \widetilde g^{\perp\perp}_{\mu\alpha}(0)\ =\ 0\ ,
\end{eqnarray}
that means the equivalence of both sets of operators.

In terms of the considered operators the spin factor,
\beq
S_{\mu\alpha}=\left(k_1-k_2\right)_{\mu}\left(k_1+k'_1\right)_{\alpha}\ ,
\eeq
is represented as:
\begin{eqnarray}
\label{xi.s.1}
\rm{Expansion \,\,\, I:} &&S_{\mu\alpha}\ =\ \xi_T(
s, s',q^2)\, g^{\perp\perp}_{\mu\alpha} +\xi_L( s, s',q^2)\,
L_{\mu\alpha}
\\
\nn
&&\xi_T( s, s', q^2)\ =\
2\left(m^2+\frac{ s s' q^2}
{\lambda( s, s', q^2)}\right),
\\
\nn
&& \xi_L( s, s', q^2)\ =\
\frac{2( s+ s'- q^2)( s- s'- q^2)
( s- s'+ q^2)}
{\lambda( s, s', q^2)}\ .
\\
\rm{Expansion \,\,\, II:}
&&S_{\mu\alpha}\ =\
\widetilde\xi_T( s, s', q^2)\,
\widetilde g^{\perp\perp}_{\mu\alpha}
+\widetilde\xi_L( s, s', q^2)\,
L_{\mu\alpha}
\\
\nn
&&\widetilde\xi_T( s, s', q^2)\ =\
2\left(m^2+\frac{ s s' q^2}
{\lambda( s, s', q^2)}\right),
\\
\nn
&& \widetilde\xi_L( s, s', q^2)\ =\
2\left(4m^2+
\frac{4 s s' q^2+( s+ s'- q^2)
( s- s'- q^2)( s- s'+ q^2)}
{\lambda( s, s', q^2)}
\right)\
\end{eqnarray}
with function ${\lambda( s, s', q^2)}$ given in (\ref{32});
the calculation of coefficients is carried out in Appendix A.
As one can see,
\beq
\xi_T( s, s', q^2)\ =\
\widetilde\xi_T( s, s', q^2),
\eeq
and this is important for the discussion.

\subsection{The amplitude of Fig. 3a.}

The whole amplitude is a sum of amplitudes of three types represented
by Fig. 3a,b,c. Let us start with the diagram of
Fig. 3a, which contains
double pole term. Corresponding amplitude is written as follows:
\beq
A^{(1\to0)}_\alpha(s,s',q^2)=(p_1-p_2)_\mu
\, \frac{g_1(s)}{1-B_1(s)}\,
\Gamma^{(1\to0)}_{\mu\alpha}(P,P';q)\,
\frac{g_0(s')}{1-B_0(s')}\ ,
\label{3.9}
\eeq
The functions $g_1(s)$  and
$g_0(s')$ are vertices of vector and scalar states;
$\Gamma_{\mu\alpha}$ is the three-point amplitude of Fig. 4a,
for which the expressions for different choices of operators read:
\begin{eqnarray}
\rm{Expansion\,\, I:}
&&\Gamma^{(1\to0)}_{\mu\alpha}(P,P';q) = g^{\perp\perp}_{\mu\alpha }
F^{(1\to0)}_T(s,s',q^2)+L_{\mu\alpha}F^{(1\to0)}_L(s,s',q^2) ,
\nn
\label{3.10.I}
\\
\rm{Expansion\,\, II:}
&&\Gamma^{(1\to0)}_{\mu\alpha}(P,P';q)=
\widetilde g^{\perp\perp}_{\mu\alpha }\widetilde F^{(1\to0)}_T(s,s',q^2)
+L_{\mu\alpha}\widetilde F^{(1\to0)}_L(s,s',q^2) .
\end{eqnarray}
Here $F^{(1\to0)}_T$, $F^{(1\to0)}_L$ and $\widetilde F^{(1\to0)}_T$,
$\widetilde F^{(1\to0)}_L$
are the form factors, for which the dispersion relations can be written
in similarly to what was described in the previous Section.

The form factor $F_i^{(1\to0)}(s,s',q^2)$ in Expansion I
reads:
\beq \label{3.15}
F^{(1\to0)}_i(s,s',q^2)\ =\ f^{(1\to0)}_i(s,s',q^2)+
\int\limits^\infty_{4m^2}\frac{d\tilde s}\pi\,\frac{d\tilde s\,'}\pi
\frac{disc_{\tilde s} disc_{\tilde s\,'}\,
F^{(1\to0)}_i(\tilde s,\tilde s\,',q^2)}
{(\tilde s-s-i0)(\tilde s\,'-s'-i0)} ,\quad i=T,L \ .
\eeq
Here $f^{(1\to0)}_i(s,s',q^2)$ is the subtraction term,
and the double spectral density is:
\begin{eqnarray}
&& disc_{\tilde s} disc_{\tilde s\,'}
F^{(1\to0)}_i(\tilde s,\tilde s\,',q^2)\
=\ \xi_i(\tilde s,\tilde s\,',q^2)\, g_1(\tilde s) g_0(\tilde s\,')\,
d\Phi_{tr}\left(\tilde P,\tilde P';k_1,k'_1,k_2\right)\ .
\label{3.16}
\end{eqnarray}
The form factor $\widetilde F_i^{(1\to0)}(s,s',q^2)$ in Expansion
II is written similarly to (\ref{3.15})  but
with differently defined double spectral density
given by  (\ref{3.16}):
one should substitute
$\xi_i(\tilde s,\tilde s',q^2)\to \widetilde\xi_i(\tilde s,\tilde
s',q^2)$.

Using dispersion relations for
the form factors, one can investigate these
two variants for any  $q^2$. But the subject of our interest is the case
of $q^2\to0$, so concentrate our attention just here.
At $q^2\to0$ the amplitude in     Expansion I takes the form:
\begin{eqnarray}
&&\Gamma^{(1\to0)}_{\mu\alpha}(s,s';q^2\to 0)\  =
\nn
\\
=&&\left[g_{\mu\alpha} +\frac{4s}{(s-s')^2}\,q_{\mu}q_\alpha
-\frac2{s-s'}\, \left(P_{\mu}q_\alpha +q_{\mu} P_\alpha\right)\right]
F^{(1\to0)}_T  (s,s',0)
+
\nn
\\
+&&\left[\frac s{(s-s')^2}\,q_{\mu}q_\alpha-\frac1{2(s-s')}\,
P_{\mu}q_\alpha\right] F^{(1\to0)}_L(s,s',0)\ =
\nn
\\
=&&\left[g_{\mu\alpha}-\frac2{s-s'}\,q_{\mu} P_\alpha\right]
F^{(1\to0)}_T  (s,s',0)
+
\nn
\\
+&&\left[\frac s{(s-s')^2}\,q_{\mu}q_\alpha-\frac1{2(s-s')}\,
P_{\mu}q_\alpha\right]
\left(F^{(1\to0)}_L(s,s',0)+4\, F^{(1\to0)}_T  (s,s',0)\right)
\ .
\label{3.21.I}
\end{eqnarray}
Hence one has:
\begin{eqnarray}
&&\widetilde F^{(1\to0)}_T(s,s',0)\ =\ F^{(1\to0)}_T(s,s',0)\ ,
\label{sootn.1}
\nn \\
&&\widetilde F^{(1\to0)}_L(s,s',0)\ =\
F^{(1\to0)}_L(s,s',0)+4\, F^{(1\to0)}_T(s,s',0)\ .
\label{sootn}
\end{eqnarray}
The calculation of three-point form factors in the limit
$q^2\to 0$ is given in Appendix B for both expansion types,
it is similar to the calculation of the three-point amplitude performed
in previous Section for the transition
$S\to\gamma S$.
Here we present the case of Expansion I only, for
 the amplitudes of Expansion II
are defined by (\ref{sootn.1}).  One gets:
\begin{eqnarray}
\label{3.24.2.I}
\rm{Expansion\,\, I:}
&&F^{(1\to0)}_i(s,s',0)= f^{(1\to0)}_i(s,s',0)+
\frac{B^{(1\to0)}_i(s)-B^{(1\to0)}_i(s')}{s-s'}, \,
 i=T,L .
\end{eqnarray}
where the loop diagram $B^{(1\to0)}_i(s)$ is equal to:
\begin{eqnarray}
\label{B.zeta}
&& B^{(1\to0)}_i(s)\ =\ \int \limits^{\infty}_{4m^2}\frac{d\tilde s}\pi\
\frac{g_1(\tilde s)g_0(\tilde s)}{\tilde s-s}\ \rho(\tilde s)\,
\zeta_i(\tilde s)\ ,
\nn
\\
&&\zeta_T(s)\ =\ 2m^2\sqrt{\frac{s}{s-4m^2}}
\ln\frac{1+\sqrt{(s-4m^2)/s}}{1-\sqrt{(s-4m^2)/s}}-s\ ,
\nn
\\
&& \zeta_L(s)\ =\ 4s\left[\sqrt{\frac s{s-4m^2}}
\ln\frac{1+\sqrt{(s-4m^2)/s}}{1-\sqrt{(s-4m^2)/s}}-2\right]\ .
\end{eqnarray}
Here we come to a clue result of our study: the
form factor related to the transverse component (\ref{sootn.1})
does not depend on the choice of expansion operators.
The choice of the expansion results in the definition of
$F^{(1\to0)}_L(s,s',0)$, but the amplitude
$A^{(1\to0)}_{L\,\,\mu\alpha}$, in its turn, does not contribute to
cross sections of physical processes with real photons because
$A^{(1\to0)}_{L\,\,\mu\alpha}\,\epsilon^{(\gamma)}_\alpha=0$.

Summing up, we conclude that in the limit $q^2\to0$  the
amplitude of Fig. 3a-type diagrams is determined unambiguously:
\beq
\label{3.31a}
A^{(1\to0)}_\alpha(s,s';0)\ =\ (p_1-p_2)_\mu
\,\,\frac{g_1(s)}{1-B_1(s)}\,\,
\Gamma^{(1\to0)}_{\mu\alpha}(s,s';0)\,\,\frac{g_0(s')}{1-B_0(s')}\ ,
\eeq
where $\Gamma^{(1\to0)}_{\mu\alpha}(s,s';0)$ is given by (\ref{3.21.I}).

\subsection{The amplitudes for the processes of Fig. 3b,c.}

Furthermore, consider the diagram of Fig. 3b, when the photon
interacts with constituents in the initial state. Corresponding
amplitude for the diagrams of such a type is given by (\ref{lt.tot});
recall that $\Gamma^{(\to0)}_\alpha(P,P';q)$ is a function represented
diagrammatically by Fig. 4b. By studying the transitions
$V\to\gamma S$, one
needs to single out the $P$-wave component in the initial state of the
pole amplitude
of Fig. 4b.  In Appendix C the expansion of pole diagram in
partial waves is presented in more detail. After singling out the
$P$-wave, the amplitude
$\Gamma^{(\to0)}_\alpha(P,P';q)$ turns into $\Gamma^{(\rm
P\to0)}_{\mu\alpha}(P,P';q)$:
\beq \Gamma^{(\rm
P\to0)}_{\mu\alpha}(P,P';q)\ =\ \frac{3}{p^2}\,
\int\frac{d\Phi_2(P;p_1,p_2)}{\rho(s)}\,
(p_1-p_2)_{\mu}\,
\Gamma^{(\to0)}_\alpha(P,P';q)\ ,
\eeq
where $p^2=(p_1-p_2)^2=4m^2-s$.

Therefore, the amplitude for diagrams with the
$P$-wave initial state takes the form:
\beq \label{3.32}
A^{(\rm P\to0)}_\alpha(P,P';q)\ =\
(p_1-p_2)_\mu\,\Gamma^{(\rm P\to0)}_{\mu\alpha}\ (P,P';q)\
\frac{g_0(s')}{1-B_0(s')}\ ,
\eeq
Now we can perform the expansion similar to what has been done in
Section 3.1, namely:
\begin{eqnarray}
\label{3.33.I}
\rm{Expansion\,\, I:}
&&\Gamma^{(\rm P\to0)}_{\mu\alpha}(P,P';q) =
g^{\perp\perp}_{\mu\alpha} F^{(\rm P\to0)}_T
(s,s',q^2)+L_{\mu\alpha} F^{(\rm P\to0)}_L(s,s',q^2) ,
\nn \\
\label{3.33.II}
\rm{Expansion\,\, II:}
&&\Gamma^{(\rm P\to0)}_{\mu\alpha}(P,P';q) =
\widetilde g^{\perp\perp}_{\mu\alpha}
\widetilde F^{(\rm P\to0)}_T(s,s',q^2)
+L_{\mu\alpha} \widetilde F^{(\rm P\to0)}_L(s,s',q^2) .
\end{eqnarray}
The form factors $F^{(\rm P\to0)}_T$, $F^{(\rm P\to0)}_L$ and
$\widetilde F^{(\rm P\to0)}_T$, $\widetilde F^{(\rm P\to0)}_L$
entering this expression may be found in the same way as for
$S\to\gamma S$, see also Appendix C.
For Expansion I we have:
\begin{eqnarray}
&&F^{(\rm P\to0)}_i(s,s',q^2)\ = \
f^{(\rm P\to0)}_i(s,s',q^2)\ +
\nn
\\
&&+\frac{3}{p^2}\int\limits^\infty_{4m^2}
\frac{d\tilde s\,'}\pi\,\frac{g_0(\tilde s\,')}{\tilde s\,'-s'}\,
\xi_i(s,\tilde s\,',q^2)
d\Phi_{tr}(P,\tilde P\,';k_1 , k'_1, k_2)\ ,\qquad
i=T,L\ ,
\label{3.40.I}
\end{eqnarray}
with factors $\xi_i(s,\tilde s',q^2)$ determined by the formula
(\ref{xi.s.1}). The formulae for
$\widetilde F^{(\rm P\to0)}_i(s,s',q^2)$ are given by an equation
similar to (\ref{3.40.I}), with the substitution $\xi_i(s,\tilde
s',q^2)\to\widetilde\xi_i(s,\tilde s',q^2)$.

In the limit $q^2\to0$, the amplitude (\ref{3.33.I}) reads:
\begin{eqnarray}
&&\Gamma^{(\rm P\to0)}_{\mu\alpha}(s,s',q^2\to0)\  =
\nn
\\
=&&\left[g_{\mu\alpha} +\frac{4s}{(s-s')^2}\,q_{\mu}q_\alpha
-\frac2{s-s'}\, \left(P_{\mu}q_\alpha+q_\mu P_\alpha \right)\right]
F^{(\rm P\to0)}_T  (s,s',0)
+
\nn
\\
+&&\left[\frac s{(s-s')^2}\,q_{\mu}q_\alpha-\frac1{2(s-s')}\,
P_{\mu}q_\alpha\right] F^{(\rm P\to0)}_L(s,s',0)\ =
\nn
\\
=&&\left[g_{\mu\alpha}-\frac2{s-s'}\,q_\mu P_\alpha\right]
F^{(\rm P\to0)}_T  (s,s',0)
+
\nn
\\
+&&\left[\frac s{(s-s')^2}\,q_{\mu}q_\alpha-\frac1{2(s-s')}\,
P_{\mu}q_\alpha\right]
\left(F^{(\rm P\to0)}_L(s,s',0)+4\, F^{(\rm P\to0)}_T  (s,s',0)\right)\ .
\label{3.21.I.lt}
\end{eqnarray}
Hence the form factors at Expansions I and II are related to each other
as:
\begin{eqnarray}
&&\widetilde F^{(\rm P\to0)}_T(s,s',0)\ =\ F^{(\rm P\to0)}_T(s,s',0)\ ,
\nn
\\
&&\widetilde F^{(\rm P\to0)}_L(s,s',0)\ =\
F^{(\rm P\to0)}_L(s,s',0)+4\, F^{(\rm P\to0)}_T(s,s',0)\ ,
\label{sootn1}
\end{eqnarray}
that is similar to  (\ref{sootn.1}).
The calculation of form factors in the limit $q^2\to0$ is performed in
Appendix B for Expansion I. We have:
\begin{eqnarray}
\label{3.43.I}
\rm{Expansion\,\,\, I:} &&F^{(\rm P\to0)}_i(s,s',0)\ =\
f^{(\rm P\to0)}_i(s,s',0)+ \frac{g_0(s)}{s-s'}\
\frac{3\,\zeta_i(s)}{4m^2-s}\ ,\quad  i=T,L \ ,
\end{eqnarray}
where $\zeta_T(s)$ and $\zeta_L(s)$ are
given in (\ref{B.zeta}).
Let us emphasize that the factor $\xi_i(s)/(4m^2-s)$
in (\ref{3.43.I}) is analytical
at $s=4m^2$, since $\xi_i(4m^2)=0$.

The amplitude for diagrams of Fig. 3c-type, with a separated $S$-wave
in the final state, reads as follows:
\beq
\label{3.45} A^{(1\to\rm
S)}_\alpha(P,P';q)\ =\ (p_1-p_2)_\mu \,\frac{g_1(s)}{1-B_1(s)}\,
\Gamma^{(1\to\rm S)}_{\mu\alpha}(s,s';q^2)\ ,
\eeq
where $\Gamma^{(1\to\rm S)}_{\mu\alpha}$ is the function represented
by Fig. 4c, where the separation of the
$S$-wave has been carried out.
For Expansion I, this function is written as follows:
\beq
\label{3.46.I}
{\rm Expansion\,\,\, I:}\quad
\Gamma^{(1\to\rm S)}_{\mu\alpha}(P,P';q)\
=\ g^{\perp\perp}_{\mu\alpha} F^{(1\to\rm S)}_T
(s,s',q^2)+L_{\mu\alpha} F^{(1\to\rm S)}_L(s,s',q^2)\ .
\eeq
The calculation of $F^{(1\to\rm S)}_i$ and $\widetilde
F^{(1\to\rm S)}_i$ can be done quite similarly to a former case. As a
result, we have the following expressions for the form factors:
\beq
{\rm Expansion\,\,\, I:}\quad F^{(1\to\rm
S)}_{T,L}(s,s',q^2)=f^{(1\to\rm
S)}_{T,L}(s,s',q^2)+\int\limits^\infty_{4m^2} \frac{d\tilde s}\pi\
\frac{g_1(\tilde s)}{\tilde s-s}\,
\xi_{T,L}(\tilde s,s',q^2)d\Phi_2(\tilde
P;k_1,k_2)\ ,
\label{3.48.I}
\eeq
In the limit $q^2\to0$,
which is just a subject of our investigation, we have:
\beq
{\rm Expansion\,\,\, I:} \quad F^{(1\to\rm
S)}_{T,L}(s,s',0)\ =\ f^{(1\to\rm S)}_{T,L}(s,s',0)
+\frac{g_1(s')}{s'-s}\zeta_i(s')\  .
\label{3.49.I}
\eeq
For Expansion I, the final amplitude at $q^2\to0$ reads:
\begin{eqnarray}
\label{3.50.I}
&& A^{(1\to\rm S)}_\alpha(s,s';0)
\ =\ (p_1-p_2)_\mu
\,\frac{g_1(s)}{1-B_1(s)}\,
\times
\\
&&\times
\left[\left(g_{\mu\alpha}+\frac{4s}{(s-s')^2}\,q_{\mu} q_\alpha
-\frac2{s-s'}\left(P_{\mu}q_\alpha+q_{\mu} P_\alpha\right)\right)
\right .\times
\nn
\\
&&\times
\left(f^{(1\to\rm S)}_T(s,s',0)
+\frac{g_1(s')}{s'-s}\zeta_T(s')\right)+
\nn
\\
+&&\left . \left(\frac{s}{(s-s')^2}q_{\mu} q_\alpha
-\frac{1}{2(s-s')}\,P_{\mu}q_\alpha\right)
\left(f^{(1\to\rm S)}_L(s,s',0)
+\frac{g_1(s')}{s'-s}\zeta_L(s')\right)\right]\ .
\nn
\end{eqnarray}
It can be easily rewritten in the form of Expansion II:
\begin{eqnarray}
&& A^{(1\to\rm S)}_\alpha(s,s';0)\ =\ (p_1-p_2)_\mu
\,\frac{g_1(s)}{1-B_1(s)}\,
\times
\\
&&\times
\left[\left(g_{\mu\alpha}-\frac2{s-s'}\,q_{\mu} P_\alpha\right)
\left(f^{(1\to\rm S)}_T(s,s',0)
+\frac{g_1(s')}{s'-s}\zeta_T(s')\right)+
\right .
\nn
\\
+&&\left(\frac{s}{(s-s')^2}\,q_{\mu} q_\alpha
-\frac{1}{2(s-s')}\,P_{\mu} q_\alpha\right)\times
\nn
\\
&&\times\left .\left(f^{(1\to\rm S)}_L(s,s',0)+
4\,f^{(1\to\rm S)}_T(s,s',0)
+\frac{g_1(s')}{s'-s}\left(\zeta_L(s')+
4\,\zeta_T(s')\right)\right)\right]\ .
\nn
\end{eqnarray}

\subsection{Analytical properties of the amplitude $V\to\gamma S$}

Now let us turn to the whole amplitude, which is the sum of processes
shown in Fig. 3a,b,c, and investigate its analytical properties in the
limit $q^2\to0$.

For the two representations of the amplitude corresponding
to Expansions I and II, one has:
\begin{eqnarray}
\label{ampl.p.1}
&& A^{(connected)}_{\mu\alpha}(s,s';0)\ =\
\frac{g_1(s)}{1-B_1(s)}
\times
\\
\nn
&&\times\left[\left(g_{\mu\alpha}+\frac{4s}{(s-s')^2}\,q_{\mu} q_\alpha
-\frac2{s-s'}\left(P_{\mu} q_\alpha+q_{\mu} P_\alpha\right)\right)\,
T(s,s',0)+
\right .
\\
\nn
&&+\left .\left(\frac{s}{(s-s')^2}\,q_{\mu} q_\alpha
-\frac{1}{2(s-s')}\,P_{\mu} q_\alpha\right)\,L(s,s',0)\right]
\frac{g_0(s')}{1-B_0(s')}\ =\
\\
\nn
=&&\frac{g_1(s)}{1-B_1(s)}
\left[\left(g_{\mu\alpha}-\frac2{s-s'}\,q_{\mu} P_\alpha\right)\,
T(s,s',0)+
\right .
\\
\nn
&&+\left .\left(\frac{s}{(s-s')^2}\,q_{\mu} q_\alpha
-\frac{1}{2(s-s')}\,P_{\mu} q_\alpha\right)\,
\Biggl(L(s,s',0)+4\,T(s,s',0)\Biggr)\right]
\frac{g_0(s')}{1-B_0(s')}\ ,
\end{eqnarray}
where
\begin{eqnarray}
\label{T.1}
T(s,s',0)\ =&&
\frac{1-B_1(s)}{g_1(s)}\,f^{(\rm P\to0)}_T(s,s',0)
+\frac{1-B_1(s)}{g_1(s)}\,\frac{g_0(s)}{s-s'}\frac{3\,\zeta_T(s)}{4m^2-s}+
\\
\nn
&&+f^{(1\to0)}_T(s,s',0)+\frac{B^{(1\to0)}_T(s)-B^{(1\to0)}_T(s')}{s-s'}+
\\
\nn
&&+f^{(1\to\rm S)}_T(s,s',0)\,\frac{1-B_0(s')}{g_0(s')} +
\frac{g_1(s')}{s'-s}\,\zeta_T(s')\,\frac{1-B_0(s')}{g_0(s')}
\end{eqnarray}
and
\begin{eqnarray}
\label{L.1}
L(s,s',0)\ =&&
\frac{1-B_1(s)}{g_1(s)}\,f^{(\rm P\to0)}_L(s,s',0)
+\frac{1-B_1(s)}{g_1(s)}\,\frac{g_0(s)}{s-s'}\frac{3\,\zeta_L(s)}{4m^2-s}+
\\
\nn
&&+f^{(1\to0)}_L(s,s',0)+\frac{B^{(1\to0)}_L(s)-B^{(1\to0)}_L(s')}{s-s'}+
\\
\nn
&&+f^{(1\to\rm S)}_L(s,s',0)\,\frac{1-B_0(s')}{g_0(s')} +
\frac{g_1(s')}{s'-s}\,\zeta_L(s')\,\frac{1-B_0(s')}{g_0(s')}\ .
\end{eqnarray}
Looking on the last equation in (\ref{ampl.p.1}), which corresponds to
Expansion II, we conclude that the analyticity of the amplitude
 $A^{(connected)}_{\mu\alpha}(s,s',0)$ requires
 the following ultimate expressions be fulfilled at $s\to s'$:
\begin{eqnarray}
\Bigl[T(s,s',0)\Bigr]_{s\to s'}\ =\ 0\ ,
\label{usl1}
\end{eqnarray}
and
\begin{eqnarray}
\label{usl2}
&&\Bigl[L(s,s',0)+4\,T(s,s',0)\Bigr]_{s\to s'}\ =\ 0\ ,
\\
\nn
&&\Bigl[\frac{L(s,s',0)+4\,T(s,s',0)}{s-s'}\Bigr]_{s\to s'}\ =\ 0\ .
\end{eqnarray}
After satisfying the requirements given by (\ref{usl1}) and
(\ref{usl2}),
the point $s=s'$ is not singular for
$A^{(connected)}_{\mu\alpha}(s,s',0)$.

First, consider the condition (\ref{usl1}) for the transverse amplitude
$T(s,s',0)$. This amplitude is defined in (\ref{T.1}), it contains pole
singularities $1/(s-s')$, which are due to both
$A^{(\rm P\to0)}_{\mu\alpha}(s,s',0)$  and
$A^{(1\to\rm S)}_{\mu\alpha}(s,s',0)$,
see (\ref{3.49.I}). These singularities should be cancelled by
similar singular points, correspondingly, in $f_T^{(\rm
P\to0)}(s,s',0)$ and $f_T^{(1\to\rm S)}(s,s',0)$. Namely, at $s\to s'$
we should deal with finite limits for:
\begin{eqnarray} &&f_T^{(\rm
P\to0)}(s,s',0)+\frac{g_0(s)}{s-s'}\,\frac{3\,\xi_T(s)}{4m^2-s}\equiv
l_T(s,s',0)\ , \\ &&f_T^{(1\to\rm
S)}(s,s',0)-\frac{g_1(s')}{s-s'}\,\xi_T(s')\equiv r_T(s,s',0)\ .
\end{eqnarray}
Therefore, $l_T(s,s,0)$ and $r_T(s,s,0)$ should be analytical
functions of $s$.  In terms of $l_T(s,s,0)$ and
$r_T(s,s,0)$, the equation (\ref{usl1}) reads:
\begin{eqnarray}
-f^{(1\to0)}_T(s,s,0)\ =\ \frac{1-B_1(s)}{g_1(s)}\,l_T(s,s,0)+
F^{(1\to0)}_T(s,s,0)+r_T(s,s,0)\,\frac{1-B_0(s)}{g_0(s)}\ .
\label{usl4}
\end{eqnarray}
Here we use the equality
\begin{eqnarray}
F^{(1\to0)}_{T}(s,s',0)\
=\ \frac{B^{(1\to0)}_T(s)-B^{(1\to0)}_T(s')}{s-s'}\ ,
\end{eqnarray}
that results in $F^{(1\to0)}_{T}(s,s,0)=dB^{(1\to0)}_T(s)/ds$.
The freedom in choosing subtraction terms makes (\ref{usl4})
being fulfilled.
Assuming that transition form factor of composite systems is defined by
double spectral integral only, the following requirement should be
imposed:
\begin{eqnarray}
f^{(1\to0)}_T(M^2_1,M^2_0,q^2)\ =\ 0\ ,
\label{usl3}
\end{eqnarray}
in particular, at $q^2=0$: $f^{(1\to0)}_T(M^2_1,M^2_0,0)=0$. We see
that the  requirement (\ref{usl3}) does not contradict the amplitude
analyticity constraint given by  (\ref{usl4}).

Likewise, using the freedom for the choice of subtraction functions,
\beq
f^{(\rm P\to0)}_L(s,s',0)\ ,\qquad
f^{(1\to0)}_L(s,s',0)\ ,\qquad f^{(1\to\rm S)}_L(s,s',0)\ ,
\eeq
one can satisfy the analyticity  constraints given by
(\ref{usl2}); still, we would not present explicitly these
 constraints, for they are rather cumbersome and do not teach us of
a something new.

\section{Quark model}

To rewrite formulae of Sections 2 and 3 for composite
fermion--antifermion systems does not provide problems.
For such  case
one needs to introduce spin variables and substitute vertices of
scalar (pseudoscalar) constituents by fermion ones:
\begin{eqnarray}
&&g_0 \ \to \ (\bar u\, u)\,g_0\ , \nn \\ &&p_{\perp\,\mu}\,g_1 \ \to \
(\bar u\gamma^{\perp}_\mu u)\,g_1\ ,
\end{eqnarray}
where $u$ is the four-spinor.
Then the consideration of fermion--antifermion composite system
$f\bar f$ remains in principle the same as for scalar (pseudoscalar)
constituents. Namely, one should consider the $(f\bar f)$
scattering amplitude of Fig. 1, and the pole of the
amplitude  $f\bar f\to f\bar f$ determines the bound state of the
$f\bar f$-system.  Its form factor
is defined by triangle graph of Fig. 2c, which is a residue in the
amplitude poles of the transition $(f\bar f)_{in}\to\gamma +(f\bar
f)_{out}$, Fig. 2b. Triangle diagram shown separately in Fig. 4a
determines  form factor of composite particle.

Still, quarks do not leave the confinement trap as free particles do,
so one cannot use for quarks the above-described scheme in a full scale.

The logic of the quark model tells us that we may treat constituent
quarks inside the confinement region as free particles, and the region,
where they are "allowed" to be free, is determined by  quark wave
function. This means that, within quark model, one can calculate the
three-point form factor amplitudes, which refer to the interaction of
photon in the intermediate state: these are the diagrams of Fig.
4a-type. The diagrams with photon interacting with incoming/outgoing
particle (Fig. 4b,c-type) should be treated using hadronic language. So
we come to a combined approach, where incoming/outgoing particles in the
processes of Fig. 3 are mesons, while constituents of the triangle
diagram are quarks. It is obvious that in such a combined approach to
the amplitude the relations (\ref{usl1}) and (\ref{usl2})  imposed by
analyticity are kept; they may be satisfied, without any problem, by
the constraints similar to  (\ref{usl4}) and (\ref{usl3}). Triangle
diagrams determined as residues in amplitude poles, see Fig. 2b, stand
for quark form factors, which are obtained in accordance with gauge
invariance and analyticity constraints.

Now, to illustrate the above statements, let us repeat the main items
of the method of singling out the quark form factor of the transition
$\phi(1020) \to \gamma f_0(980)$ from hadronic processes. We cannot use
directly the process $(q\bar q)_{incoming}\to \gamma
(q\bar q)_{outgoing}$ for the definition of quark form factor: being
rigorous we are not allowed to treat quarks as free particles at large
distances. In this way, the reaction of the type $K\bar K\to \gamma
\pi\pi$ is to be considered; then the amplitude $\phi(1020)\to \gamma
f_0(980)$ should be extracted as a double residue in the
amplitude poles of
$K\bar K \to \gamma\pi\pi$. The amplitude itself, $\phi(1020)\to \gamma
f_0(980)$, may be considered in terms of constituent quarks, namely, as
triangle diagram with constituent quarks.

The form factor amplitude for radiative transition of vector meson to
scalar one, of the type of $\phi(1020)\to\gamma f_0(980)$,
was calculated in terms of
constituent quarks  \cite{epja,rad-yf}.
The spectral integral for $F^{(1\to0)}_T(M^2_V,M^2_S,0)$ was given by
Eq.  (32) of \cite{rad-yf}, it was denoted there as
$A_{V\to\gamma S}(0)$. Up to the charge factor
$Z_{V\to\gamma S}$, which we omit here, it reads:
\begin{eqnarray}
F^{(1\to 0)}_T(M^2_V,M^2_S,0) =
\int\limits^{\infty}_{4m^2}
\frac{ds}{\pi}\Psi_V(s)\Psi_S(s)
\left[\frac{m}{4\pi}\sqrt{s(s-4m^2)}-
\frac{m^3}{2\pi}\ln\frac{\sqrt{s}+\sqrt{s-4m^2}}{\sqrt{s}-\sqrt{s-4m^2}}
\right] .
\label{4.2}
\end{eqnarray}
The quark wave functions for vector and scalar states,
$\Psi_V(s)$ and $\Psi_S(s)$, are normalized as follows:
\begin{eqnarray}
\label{4.2a}
\int\limits^{\infty}_{4m^2}
\frac{ds}{\pi}\Psi_V^2(s)
\frac{s+2m^2}{12\pi}\sqrt{\frac{s-4m^2}{s}}=1\, , \quad
\int\limits^{\infty}_{4m^2}
\frac{ds}{\pi}\Psi_S^2(s)
\frac{s-4m^2}{8\pi}\sqrt{\frac{s-4m^2}{s}}=1\, .
\end{eqnarray}
Masses of light constituent quarks $u$
and $d$ are of the order of $350$ MeV, and the strange-quark mass $\sim
500$ MeV.

The function in square brackets of the integrand (\ref{4.2}) is
positive at $s > 4m^2$, so the transition form factor
$F^{(1\to 0)}_T(M^2_V,M^2_S,0) \neq 0$ at arbitrary $M_V$ and $M_S$,
$M_V=M_S$ included, if the wave functions $\Psi_V(s)$ and
$\Psi_S(s)$ do not change sign, and just this feature (absence of
zeros in  radial wave function) is a signature of basic states
with the radial quantum number $n=1$.

One point needs special discussion, that is, the possibility to apply
spectral-integration technique, with mass-on-shell
intermediate states, to the calculation of quark diagrams of
Fig. 4a. The fact that quarks--constituents do not fly off at
large distances does not restrain directly the
calculation technique: calculations may be performed with
mass-on-shell particles in the intermediate states as well as
mass-off-shell ones, like  Feynman integrals.
Recall that in non-relativistic quantum mechanics, that is,  in
non-relativistic quark model, the particles in the intermediate states
are  mass-on-shell, that does not prevent the use of confinement models.
The flying-off of quarks at
large distances corresponds to threshold singularities of the
amplitude: in the triangle diagram of Fig. 4a  they are
at $s=4m^2$ and $s'=4m^2$.  The suppression of
contribution from large distances means the
suppression  in the momentum space  from the regions
$s\sim 4m^2$ and $s'\sim 4m^2$. Such a suppression is implemented by the
properties of vertices, or wave functions,
of composite systems. As concern the threshold
singularities, they are present in other techniques too, like
Feynman or light-cone ones.
Therefore, in all representations one should suppress
the contributions from the regions  $s\sim 4m^2$ and
$s'\sim 4m^2$. The spectral-integration technique provides us
 a possibility  to control the near-threshold contributions.

\subsection*{ Transition $^3S_1q\bar q\to
^3P_0q\bar q$ in the non-relativistic approach}

In the non-relativistic limit, formula (\ref{4.2}) turns into standard
expression of the quark model transition $^3S_1\to ^3P_0$, and
non-relativistic amplitude is obtained
by expanding
the expression in square brackets
in a series in respect to relative
quark momentum squared $k^2$, where $k^2=s/4-m^2$.
Here we present the
form factor $F^{(1\to0)}_T(M^2_V,M^2_S,0)$
in the non-relativistic approach:
this very transition is responsible for the decays
$\phi(1020)\to \gamma f_0(980)$ and $\phi(1020)\to \gamma  a_0(980)$, if
$f_0(980)$ and $a_0(980)$ are the $q\bar q$ states.

In the non-relativistic approximation,
after re-definition $\Psi_V(s)\to\psi_V(k^2)$ and
$\Psi_S(s)\to\psi_S(k^2)$,
Eq. (\ref{4.2}) turns into:
\begin{eqnarray}
F^{(1\to0)}_T(M^2_V,M^2_S,0) \to
\left[F^{(1\to0)}_T(M^2_V,M^2_S,0)\right]_{non-relativistic} =
\int\limits^{\infty}_{0}\frac{d\,k^2}{\pi}\,\psi_V(k^2)\,\psi_S(k^2)\,
\frac{8}{3\pi}\,k^3\ ,
\label{4.3}
\end{eqnarray}
Normalization conditions for wave functions
$\psi_V(k^2)$ and $\psi_S(k^2)$ should be also re-written in the
non-relativistic limit, they read:
\begin{eqnarray}
1\ =\ \int\limits^{\infty}_{0}\frac{d\,k^2}{\pi}\,\psi^2_S(k^2)\,
\frac{2k^3}{\pi m}\ ,\qquad
1\ =\ \int\limits^{\infty}_{0}\frac{d\,k^2}{\pi}\,\psi^2_V(k^2)\,
\frac{mk}{2\pi}\, .
\label{4.4}
\end{eqnarray}
For exponential parametrization of wave functions,
\begin{eqnarray}
\psi_S(k^2)=N_Se^{-b_Sk^2}\ ,\qquad \psi_V(k^2)=N_Ve^{-b_Vk^2}\ ,
\label{4.5}
\end{eqnarray}
one can easily calculate integrals (\ref{4.3}) and (\ref{4.4}).
Normalization constants are determined as
\begin{eqnarray}
N_S^2=\frac{8\sqrt 2}{3}\pi^{3/2}mb_S^{5/2}, \qquad
N_V^2=2\sqrt 2 \pi^{3/2}\frac{1}{m} b_S^{3/2}\, ,
\label{4.6}
\end{eqnarray}
and transition form factor is equal to:
\begin{eqnarray}
\left[F^{(1\to0)}_T(M^2_V,M^2_S,0)\right]_{non-relativistic} =
8\sqrt{\frac{2}{3}}\frac{b_V^{3/4}b_S^{5/4}}{(b_V+b_S)^{5/2}}\, .
\label{4.7}
\end{eqnarray}
At $b_V\simeq b_S=b$, one has
\begin{eqnarray}
\left[F^{(1\to0)}_T(M^2_V,M^2_S,0)\right]_{non-relativistic} \simeq
\frac{2}{\sqrt {3b}}\, .
\label{4.8}
\end{eqnarray}
For loosely bound systems $1/b\sim \sqrt{m\epsilon}$, where $\epsilon$
is the binding energy, so the right-hand side of Eq. (\ref{4.8})
contains the suppression factor inherent in the E1-transition.

In Fig. 5 we demonstrate the calculated form factors $F^{(1\to 0)}_T$
for the transition
$\phi(1020)\to \gamma f_0(980)$
both for
non-relativistic approximation, Eq. (\ref{4.3}), and
relativistic spectral integrals, Eq. (\ref{4.2}), with the $n\bar n$
and $s\bar s$ components. We use $b_V =10  $ GeV$^{-2}$ that
corresponds to the $\phi$-meson radius of the order of pion radius,
and for the $f_0$-meson we change the wave function slope
within the limits $2$ GeV$^{-2}
<b_S<12 $ GeV$^{-2}$, that means the change of the radius squared in the
interval $0.5R^2_\pi < R^2_{f_0} < 2R^2_\pi  $, for more detail see
\cite{epja,rad-yf}.

It is seen that form factors calculated in both relativistic and
non-relativistic approaches do not differ significantly one from
another, that makes puzzling the statement about impossibility of the
quark-model description of the reactions $\phi(1020)\to \gamma
f_0(980)$ and $\phi(1020)\to \gamma  a_0(980)$ under the assumption
about $f_0(980)$ and $a_0(980)$ being $q\bar q$ states: recall that
the data can be indeed described \cite{epja,rad-yf} by using
spectral-integral formula (\ref{4.2}).

\section{Conclusion}

The use of dispersion technique for the calculation of form factors of
composite system has
certain advantages, of which we would underline two.

(i) In the dispersion technique, or in the spectral integration
technique, the content of bound state is controlled. The interaction of
constituents with each other due to meson exchanges does not lead to the
appearance of new components in the bound state
 related to these mesons.\\

(ii) The dispersion technique, as well as spectral integration
technique, works with the energy-off-shell amplitudes, and the
particles in the intermediate states are mass-on-shell. This provides
us an easy possibility to construct spin operators, that in its turn
allows one to consider without  problems
the amplitudes for composite particles with a large spin
(see \cite{operator} for more detail).

However, as it happens often, the advantages make it
necessary to take special care about other aspects of the
approach: in the
spectral integration technique, when radiative processes are
considered, one should impose "by hand" the constraints related to
gauge invariance and analyticity. In  \cite{deut}, this problem was
considered in connection with charge form factors, for example, for the
transitions of $(S\to\gamma S)$-type, where $S$ is scalar meson.
However, more complicated radiative processes were
investigated later on \cite{ff}, such as $V\to\gamma S$
($V$ is vector meson) and
$P\to\gamma\gamma$, $S\to\gamma\gamma$, $T\to\gamma\gamma$
($P$ and  $T$ are pseudoscalar and tensor mesons). In these cases
the spin structure of amplitude is more complicated, hence more
complicated constraints for amplitudes are needed, that is related to
gauge-invariant operators which perform the moment-operator
expansion.  Although in  principle the reconstruction of analyticity
and transition form factors is analogous to that used in \cite{deut},
the statements of Ref. \cite{acha} concerning transition form factors
need certain comment.

In the present paper, we have carried out the study of
constraints owing to gauge invariance and analyticity
using as an
example the process $V\to\gamma S$.
As the
first step, we accept the mesons $V$ and $S$ to be two-particle
composite systems of scalar or pseudoscalar particles--constituents
(Section 3). Such an approach makes calculations simple, less
cumbersome, without affecting basic points.  This approach can be
generalized for fermion-antifermion system without principal complications,
 and the necessary changes are related only
to the phase space (spin factors should be included) and the form of
vertices. In this way, one can generalize the present results for
quark-antiquark systems (Section 4), i.e. one can apply the method
to the consideration of radiative decays of  $q\bar q$-mesons too.

Concerning the transition $V\to \gamma S$, we demonstrate that two
independent operators are responsible for the spin structure of this
reaction, $g^{\perp\perp}_{\mu\alpha}$ and $L_{\mu\alpha}$ given by
(\ref{1.11}) and (\ref{1.12}). The operators
$g^{\perp\perp}_{\mu\alpha}$ and $L_{\mu\alpha}$ determine
transverse and longitudinal amplitudes.
In the spectral
integration technique, the essential point is the use of operators,
which are responsible for total spin space, that is, two orthogonal
operators $g^{\perp\perp}_{\mu\alpha}$ and $L_{\mu\alpha}$, with
$g^{\perp\perp}_{\mu\alpha}L_{\mu\alpha}=0$.
At $q^2=0$ the longitudinal
operator turns into a nilpotent one,
$L_{\mu\alpha}(0)L_{\mu\alpha}(0)=0$.

Also, we  demonstrate that meson-decay form factors
are determined by residues in the poles of scattering blocks.
For example, the form factor for the decay
$\phi(1020)\to\gamma f_0(980)$ is given by the residue of the amplitude
of the reaction $\phi (1020)\to \gamma \pi\pi$, so  analytical
properties of these two amplitudes are different.
In particular, the amplitude of the transition
$\phi (1020)\to \gamma \pi\pi$ should be zero at $(M_\phi -M_{\pi\pi})
\to 0$, while for the transition $\phi(1020)\to\gamma f_0(980)$
an analogous requirement is absent for $M_\phi =M_{f_0}$. Moreover,
if $\phi(1020)$ and $f_0(980)$ are members of the basic $q\bar q$
nonets, $1^3S_1$ and $1 ^3P_0$, the transition amplitude cannot be zero
at any meson masses, including $M_\phi =M_{f_0}$. An opposite statement
is declared in \cite{acha}.

Present investigation does not confirm the use  of a unique form of the
spin operator in $S\to\gamma V$ declared in \cite{acha}:  generally,
at $q^2 \neq 0$, two independent spin operators exist related
to the transverse and longitudinal amplitudes,
$g^{\perp\perp}_{\mu\alpha}$ and $L_{\mu\alpha}$.
At $q^2 \to 0$ one
operator, $L_{\mu\alpha}(0)$, turns into nilpotent one, thus giving us a
freedom in writing spin operator for the transition with real photon.

We thanks A.V. Anisovich, Ya.I. Azimov, L.G. Dakhno, M.N.
Kobrinsky, D.I. Melikhov, V.N. Nikonov and V.A. Sarantsev
for helpful discussions of problems involved.
This work is supported by the grants RFBR  N 0102-17861 and
N 00-15-96610.

\section*{Appendix A: Coefficients
$\xi_T( s, s', q^2)$ and $\xi_L( s, s', q^2)$}

Let us calculate coefficients in the expansion
(\ref{xi.s.1}) for spin factor $S_{\mu\alpha}$.
To this aim we need the convolutions as follows:
\begin{eqnarray}
&& g_{\mu\alpha}(k_1-k_2)_{\mu}(k_1+k'_1)_\alpha\
=\ 4m^2-\frac12\,(s+s'+q^2)\ , \nn \\
&& P_{\mu}P_\alpha(k_1-k_2)_{\mu}(k_1+k'_1)_\alpha\ =\ 0\ , \nn\\
&& q_{\mu}q_\alpha(k_1-k_2)_{\mu}(k_1+k'_1)_\alpha\ =\ 0\ , \nn\\
&& P_{\mu}q_\alpha(k_1-k_2)_{\mu}(k_1+k'_1)_\alpha\ =\ 0\ ,
\label{3.18}
\\
&& q_{\mu}P_\alpha(k_1-k_2)_{\mu}(k_1+k'_1)_\alpha\ =\ \frac14
 (s+s'-q^2)(-s+s'+q^2)\ .    \nn
\end{eqnarray}
Projecting  $S_{\mu\alpha}$ on the operators
$g^{\perp\perp}_{\mu\alpha} $ and
$L_{\mu\alpha} $,
one obtains:
\begin{eqnarray}
&& \xi_T(s,s',q^2)\ \left(g^{\perp\perp}_{\mu\alpha}\right)^2\
 =\ 4\left(m^2+ \frac{ss'q^2}{\lambda(s,s,q^2)}\right) , \nn
\\
&& \xi_L(s,s',q^2)\ \left(L_{\mu\alpha}\right)^2\
=\ -\frac{q^2s}{2\lambda(s,s',q^2)}\,
\frac{(s+s'-q^2) (-s+s'+q^2)}{(s-s'+q^2)}\ .
\label{3.19}
\end{eqnarray}
As a result, we have the coefficients given in (\ref{xi.s.1}).
Taking into account that
$g^{\perp\perp}_{\mu\alpha}=\widetilde g^{\perp\perp}_{\mu\alpha}
+4L_{\mu\alpha}$,
we obtain coefficients for Expansion II,
$ \widetilde\xi_T(s,s',q^2)$
and $\widetilde\xi_L(s,s',q^2)$.

\section*{Appendix B: Form factors in the limit $q^2\to 0$}

\subsubsection*{Form factors of the three-point diagram of Fig. 4a}

Let us calculate form factors in the limit $q^2\to 0$ for the
diagram  of Fig. 4a for two variants of the expansion given by
(\ref{3.10.I}).

{\bf Expansion I:}
To get the formula (\ref{3.15}) for form factors in the limit
$q^2\to0$ let us insert new variables:
$\sigma$, $\Delta$ and $Q^2$, see (\ref{new-variable}),
and integrate over phase space, that leads at small $Q^2$
to
\beq \label{3.23}
F^{(1\to0)}_i(s,s',q^2\to0)\ =\ f^{(1\to0)}_i(s,s',q^2\to0)+
\int\limits^\infty_{4m^2} \frac{d\sigma}\pi\,
\frac{g_1(\sigma)g_0(\sigma)}{(\sigma-s)(\sigma-s')}
\int\limits^b_{-b}d\Delta
\frac{\xi_i(\sigma,\Delta,Q^2)}{16\pi\sqrt{\Delta^2+4\sigma Q^2}}\ ,
\eeq
where $b$ is determined in (\ref{b}) and
\begin{eqnarray}
&&\xi_T(\sigma,\Delta,Q^2)\ =\
2\left(m^2-\frac{Q^2\sigma^2}{\Delta^2+4\sigma Q^2}\right)\ ,
\nn
\\
&& \xi_L(\sigma,\Delta,Q^2)\ =\
4\sigma\left(\frac{\Delta^2}{\Delta^2+4\sigma Q^2}\right)\ .
\label{3.23a}
\end{eqnarray}
Integrating over $\Delta$, we obtain final formula for form
factors in Expansion I, see (\ref{3.24.2.I}) and
(\ref{B.zeta}).

{\bf Expansion II:}
Likewise, the same procedure is carried out for Expansion II, though
with other coefficients defining double discontinuity of the form
factor:
\begin{eqnarray}
&&\widetilde\xi_T(\sigma,\Delta,Q^2)\ =\ \xi_T(\sigma,\Delta,Q^2)\ ,
\nn \\ &&\widetilde\xi_L(\sigma,\Delta,Q^2)\ =\
4\left(2m^2+\frac{\sigma(\Delta^2-2\sigma Q^2)}{\Delta^2+4\sigma Q^2}\right)
\label{3.31}
\end{eqnarray}
As a result, we obtain:
$\widetilde F^{(1\to0)}_T(s,s',q^2)\ =\ F_T^{(1\to0)}(s,s',0)$ and
\begin{eqnarray}
&& \widetilde F^{(1\to0)}_L(s,s',q^2)\ =\ \widetilde f_L^{(1\to0)}(s,s',0)
+\frac{\widetilde B^{(1\to0)}_L(s)-\widetilde B^{(1\to0)}_L(s')}{s-s'}\ ,
\nn
\\
&& \widetilde B^{(1\to0)}_L(s)\ =\int \limits^{\infty}_{4m^2}\frac{d\tilde s}\pi\
\frac{g_1(\tilde s)g_0(\tilde s)}{\tilde s-s}\ \rho(\tilde s)\,
\widetilde\zeta_L(\tilde s)\ ,
\nn
\\
&& \widetilde\zeta_L(s)\
=\ 4\left[(2m^2+s)\sqrt{\frac{s}{4m^2-s}}
\ln\frac{1+\sqrt{(s-4m^2)/s}}{1-\sqrt{(s-4m^2)/s}}-3s\right]\ .
\label{3.27}
\end{eqnarray}

\subsubsection*{Two-point diagram of Fig. 4b}

For the diagram of Fig. 4b, the form factor calculations in the limit
$q^2\to 0$, see (\ref{3.40.I}), is carried out as before, by
using the variables
$\sigma$, $\Delta$ and $Q^2$ determined in (\ref{new-variable}):
\beq
F^{(\rm P\to0)}_i(s,s',q^2\to0)=f^{(\rm P\to0)}_i(s,s',q^2\to0)+
\frac{3}{p^2}\int\limits^\infty_{4m^2}\frac{d\sigma}\pi
\frac{g_0(\sigma)}{\sigma-s'}\pi\delta(\sigma-s)\int\limits^{b}_{-b}d\Delta
\frac{\xi_i(\sigma,\Delta,Q^2\to0)}{16\pi\sqrt{\Delta^2+4\sigma
Q^2}}\ ,
\label{3.42}
\eeq
where $\xi_i(s,s',q^2)$ are given by (\ref{3.23a}). Performing
calculations analogous to the above ones,
we obtain (\ref{3.43.I}).

\section*{Appendix C: Separation of the $P$-wave in the initial state of
the process of Fig. 4b}

The Feynman diagram shown in Fig. 4b reads:
\beq
\Gamma^{(\to0)}_\alpha\ =\left(p_1+k'_1\right)_\alpha\,\frac{g_0(P'^2;k'^2_1,k^2_2)}{m^2-k'^2_1}\ .
\label{ampl_lt}
\eeq
The amplitude $\Gamma^{(\to0)}_\alpha$ can be expanded in a series with
respect to orbital momenta of the initial states as follows:
\beq
\Gamma^{(\to0)}_\alpha=\Gamma^{(\rm S\to0)}_\alpha
+(p_1-p_2)_{\mu}\,\Gamma^{(\rm P\to0)}_{\mu\alpha}
+X^{(2)}_{\mu_1\mu_2}(p)\,\Gamma^{(\rm D\to0)}_{\mu_1\mu_2\alpha}+\cdots
\label{angl_ex}
\eeq
There is summation over $\mu$ and $\mu_1\mu_2$.
To find out $\Gamma^{(\rm P\to0)}_{\mu\alpha}$, one should multiply
$\Gamma^{(\to0)}_{\alpha}$ by
\beq
(p_1-p_2)_{\mu}\equiv p_\mu
\eeq
and integrate over  $d\Omega/(4\pi)$ or $\int d\Phi_2(P;p_1,p_2)/\rho(s)$.
The right-hand side of (\ref{angl_ex}) gives us:
\beq
\int\frac{d\Phi_2(P;p_1,p_2)}{\rho(s)}\,(p_1-p_2)_\mu \,(p_1-p_2)_{\mu}\,
\Gamma^{(\rm P\to0)}_{\mu\alpha}\ =\
\frac{p^2}{3}\,\Gamma^{(\rm P\to0)}_{\mu\alpha}\ ,
\label{integ}
\eeq
for one can replace in the integrand of (\ref{integ}):
\beq
(p_1-p_2)_{\mu}\,(p_1-p_2)_{\mu'}\to
q^{\perp}_{\mu\mu'}\,\frac{p^2}{3}\ ,\qquad p^2=4m^2-s\ .
\eeq
The left-hand side of (\ref{angl_ex}) with the account for
(\ref{ampl_lt}) gives us:
\beq
\int\frac{d\Phi_2(P;k_1,k_2)}{\rho(P^2)}\,(k_1-k_2)_\mu
\,(k_1+k'_1)_{\alpha}\, \frac{g_0(P'^2;k'^2_1,k^2_2)}{m^2-k'^2_1}\ .
\eeq
In the integrand we
re-denoted the momenta $p_1$ and $p_2$ as $k_1$ and
$k_2$.
To get the spectral integral,  let us calculate the
discontinuity; to this aim we consider intermediate
state as mass-on-shell:
\beq \label{3.35}
(m^2-k'^2_1)^{-1}\ \longrightarrow\ \theta(k'_{10})
\delta(m^2-k'^2_1)\ ,
\eeq
and substitute the vertex function as follows:
\beq \label{3.36}
g_0(\tilde P\,'^2;k'^2_1,k^2_2)\ \longrightarrow\ g_0(\tilde s\,')\ .
\eeq
Then we expand the spin factor of the
amplitude:
\beq \label{3.39}
(k_1-k_2)_\mu(k_1+k'_1)_\alpha\ =\ g^{\perp\perp}_{\mu\alpha}
\xi_T(s,\tilde s\,', q^2)+L_{\mu\alpha}\xi_L(s,\tilde s\,',
q^2)\ ,
\eeq
with the coefficients $\xi_T$ and $\xi_L$  given in
(\ref{xi.s.1}). As a result, the discontinuity
in the $s'$-channel reads:
\begin{eqnarray}
&& disc_{\tilde s'}\,F^{(\rm P\to0)}_{T,L}(s,s',q^2)\
=\frac{1}{\rho(s)}\,\xi_{T,L}(s,\tilde s',q^2)\,
g_0(\tilde s')\,d\Phi_{tr}(P,P';k_1,k'_1,k_2)\ .
\end{eqnarray}
So we have the
following dispersion representation for the pole diagram
of Fig. 4b:
\begin{eqnarray}
&&F^{(\rm P\to0)}_{T,L}(s,s',q^2)\ =\ f^{(\rm P\to0)}_{T,L}(s,s',q^2)\ +
\nn
\\
&&+\frac{3}{p^2}\frac{1}{\rho(s)}\int\limits^{\infty}_{4m^2}\,
\frac{d\tilde s'}{\pi}\,\frac{g_0(\tilde s')}{\tilde s'-s'}\,
\xi_{T,L}(s,\tilde s',q^2)\,d\Phi_{tr}(P,P';k_1,k'_1,k_2)\ ,
\label{ff.ap.lt}
\end{eqnarray}
that gives (\ref{3.40.I}).

\newpage

\begin{figure}
\centerline{\epsfig{file=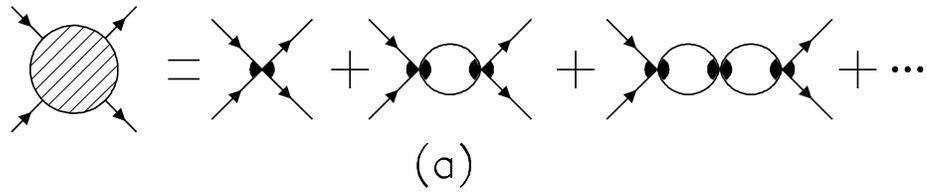,width=15cm}}
\centerline{\epsfig{file=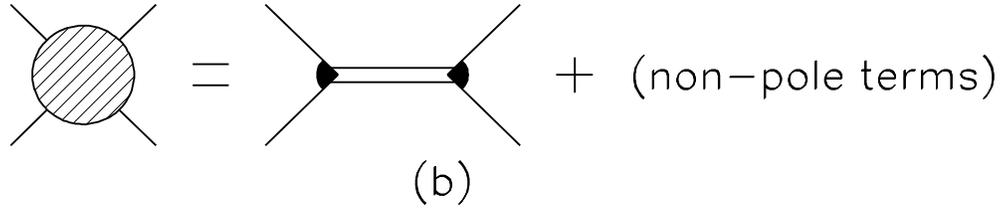,width=15cm}}
\caption{ a)
Representation of partial scattering amplitude as a set of the
dispersion loop diagrams.
b) Partial amplitude at the energies close to the mass of a bound state:
the main contribution is given by the pole at  $s=M^2$.}
\end{figure}

\begin{figure}
\centerline{\epsfig{file=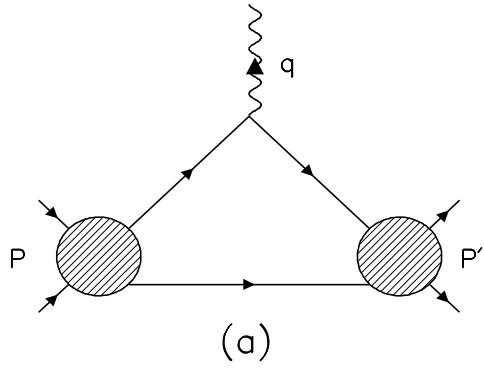,width=8cm}}
\centerline{\epsfig{file=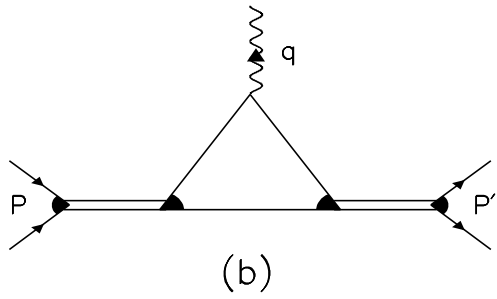,width=8cm}}
\centerline{\epsfig{file=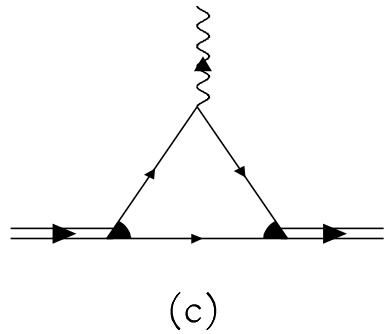,width=8cm}}
\caption{Diagrams determining  form factor of the composite system:
a) the photon emission by interacting constituents;
b) the diagram of Fig. 2a near pole corresponding to bound
states; c) form factor of bound state determined as a residue of
the amplitude of Fig. 2b in the poles at $s=M^2$ and $s'=M'^2$. }
\end{figure}

\begin{figure}
\centerline{\epsfig{file=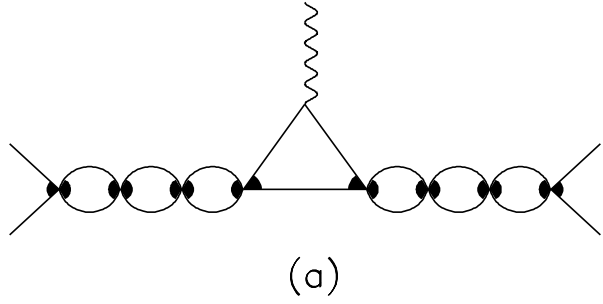,width=9cm}}
\centerline{\epsfig{file=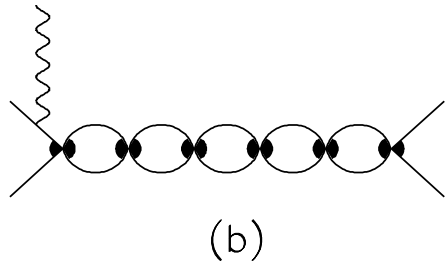,width=7cm}}
\centerline{\epsfig{file=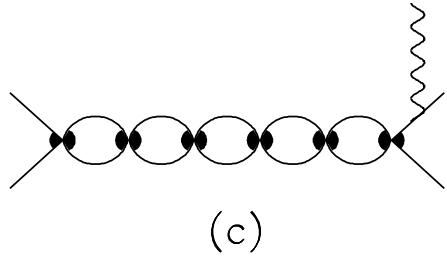,width=7cm}}
\centerline{\epsfig{file=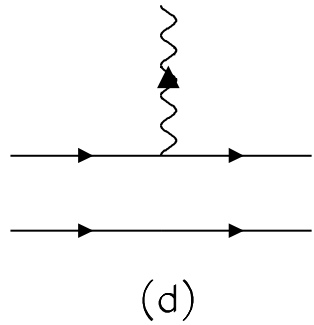,width=6cm}}
\caption{a) The process shown diagrammatically in Fig. 2a in terms of
the dispersion loop diagrams; b,c) diagrams with the photon interaction
in the initial and final states; d)  emission of photon by
non-interacting constituents.}
\end{figure}

\begin{figure}
\centerline{\epsfig{file=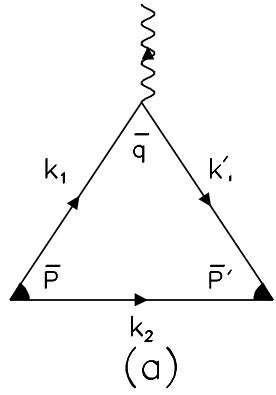,width=7cm}}
\centerline{\epsfig{file=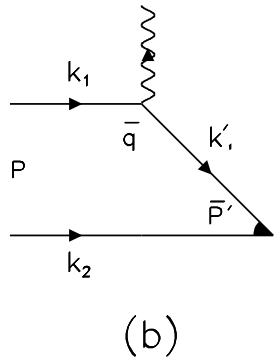,width=7cm}}
\centerline{\epsfig{file=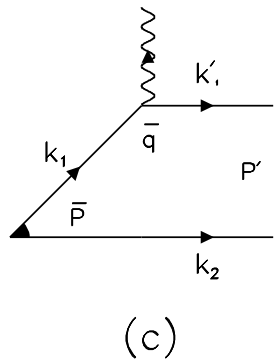,width=7cm}}
\caption{a) Three-point form factor amplitude.
The blocks for the emission of photon in the initial
(b) and final (c) states.}
\end{figure}

\begin{figure}
\centerline{\epsfig{file=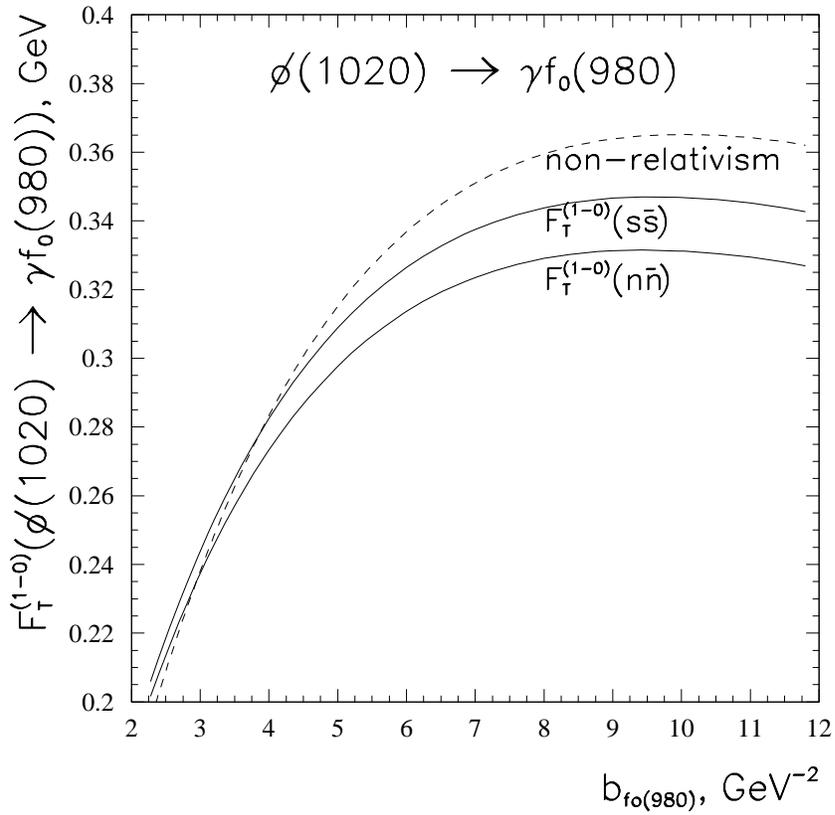,width=12cm}}
\caption{Form factor $F_T^{1\to 0}(M^2_V , M^2_S)$ for the
decay $\phi (1020) \to \gamma f_0(980)$
calculated in relativistic (for $n\bar n$
and $s\bar s$ components) and non-relativistic approaches:
solid and dashed lines.}
\end{figure}

\end{document}